\documentclass[10pt,journal]{IEEEtran} %

\ifCLASSOPTIONcompsoc
  \usepackage{cite}
\else
  \usepackage{cite}
\fi

\ifCLASSINFOpdf
  \usepackage[pdftex]{graphicx}
\else
\fi

\usepackage{amsmath}
\usepackage{amsthm}

\usepackage{algorithm}
\usepackage{algorithmicx}
\usepackage{algpseudocode}

\algnewcommand\algorithmicforeach{\textbf{foreach}}
\algdef{S}[FOR]{ForEach}[1]{\algorithmicforeach\ #1\ \algorithmicdo}

\algnewcommand\Or{\textbf{or} }
\algnewcommand\AAA{\textbf{and} }
\algnewcommand\Continue{\textbf{continue}}

\ifCLASSOPTIONcompsoc
\else
\fi
\usepackage{subcaption}

\usepackage{listings}
\usepackage{enumitem}

\usepackage{textcomp}
\usepackage{xcolor}
\usepackage{amsmath}
\usepackage{caption}
\usepackage{booktabs}
\usepackage{bbm}
\usepackage{mathtools}
\usepackage{color}
\usepackage{dsfont}
\usepackage{amssymb} %

\usepackage{hyperref}  %

\def\BibTeX{{\rm B\kern-.05em{\sc i\kern-.025em b}\kern-.08em
		T\kern-.1667em\lower.7ex\hbox{E}\kern-.125emX}}

\newcommand{\todo}[1]{{\color{red}#1}} %
\newcommand{\ignore}[1]{}

\begin{document}
\title{Quick and Reliable LoRa Physical-layer Data Aggregation through Multi-Packet Reception}

\author{Lizhao~You,~
        Zhirong~Tang, Pengbo~Wang, Guanglong~Pang, Zhaorui~Wang,~Haipeng Dai,~
        and~Liqun~Fu}
        
\author{Lizhao~You,
        Zhirong~Tang,
        Pengbo~Wang,
        Zhaorui~Wang,
        Haipeng Dai,
        and~Liqun~Fu%
\IEEEcompsocitemizethanks{\IEEEcompsocthanksitem L. You, Z. Tang, P. Wang, and L. Fu are with the Department of Information and Communication Engineering, School of Informatics, Xiamen University, China. E-mails: \{lizhaoyou,~liqun\}@xmu.edu.cn, \{ttangzr,~pbxmu\}@stu.xmu.edu.cn.
	\IEEEcompsocthanksitem  Z. Wang is with the School of Science and Engineering, The Chinese University of Hong Kong, Shenzhen, China. E-mail: wangzhaorui@cuhk.edu.cn. \IEEEcompsocthanksitem H. Dai is with the State Key Laboratory for Novel Software Technology, Nanjing University, China. E-mail: haipengdai@nju.edu.cn. %
}%
}

\IEEEtitleabstractindextext{
\begin{abstract}
This paper presents a Long Range (LoRa) physical-layer data aggregation system (LoRaPDA) that aggregates data (e.g., sum, average, min, max) directly in the physical layer. In particular, after coordinating a few nodes to transmit their data simultaneously, the gateway leverages a new multi-packet reception (MPR) approach to compute aggregate data from the phase-asynchronous superimposed signal. Different from the analog approach which requires additional power synchronization and phase synchronization, our MRP-based digital approach is compatible with commercial LoRa nodes and is more reliable. Different from traditional MPR approaches that are designed for the collision decoding scenario, our new MPR approach allows simultaneous transmissions with small packet arrival time offsets, and addresses a new co-located peak problem through the following components: 1) an improved channel and offset estimation algorithm that enables accurate phase tracking in each symbol, 2) a new symbol demodulation algorithm that finds the maximum likelihood sequence of nodes' data, and 3) a soft-decision packet decoding algorithm that utilizes the likelihoods of several sequences to improve decoding performance. 
Trace-driven simulation results show that the symbol demodulation algorithm outperforms the state-of-the-art MPR decoder by 5.3$\times$ in terms of physical-layer throughput, and the soft decoder is more robust to unavoidable adverse phase misalignment and estimation error in practice.
Moreover, LoRaPDA outperforms the state-of-the-art MPR scheme by at least 2.1$\times$ for all SNRs in terms of network throughput, demonstrating quick and reliable data aggregation.%
\end{abstract}

}

\maketitle

\IEEEdisplaynontitleabstractindextext

\IEEEpeerreviewmaketitle

\section{Introduction}\label{sec:intro}

\IEEEPARstart{A}s a representative Low-Power Wide Area Network (LPWAN) technology, LoRa has been widely used in many Internets of Things (IoT) applications such as smart cities, remote sensing, and environment  monitoring\cite{li2022lora}. In this paper, we focus on a particular data aggregation query application, where the LoRa gateway aims to collect an aggregate function (e.g., sum, average, min, max) of LoRa sensor nodes' measurement data within a short time \cite{sanchez2014smartsantander,talavera2017review,gadre2020quick}. 
Commercial off-the-shelf (COTS) LoRa nodes that adopt the simple ALOHA MAC protocol without carrier sensing is inefficient. The collisions can be very severe with a large number of nodes, and re-transmissions would be quite common \cite{adelantado2017understanding}. Moreover, due to the extremely low data rate adopted by LoRa, the packet transmission time is long, and the overall data aggregation time will be longer, further challenging quick data aggregation.

\begin{figure}
	\centering
	\includegraphics[width=0.4\textwidth]{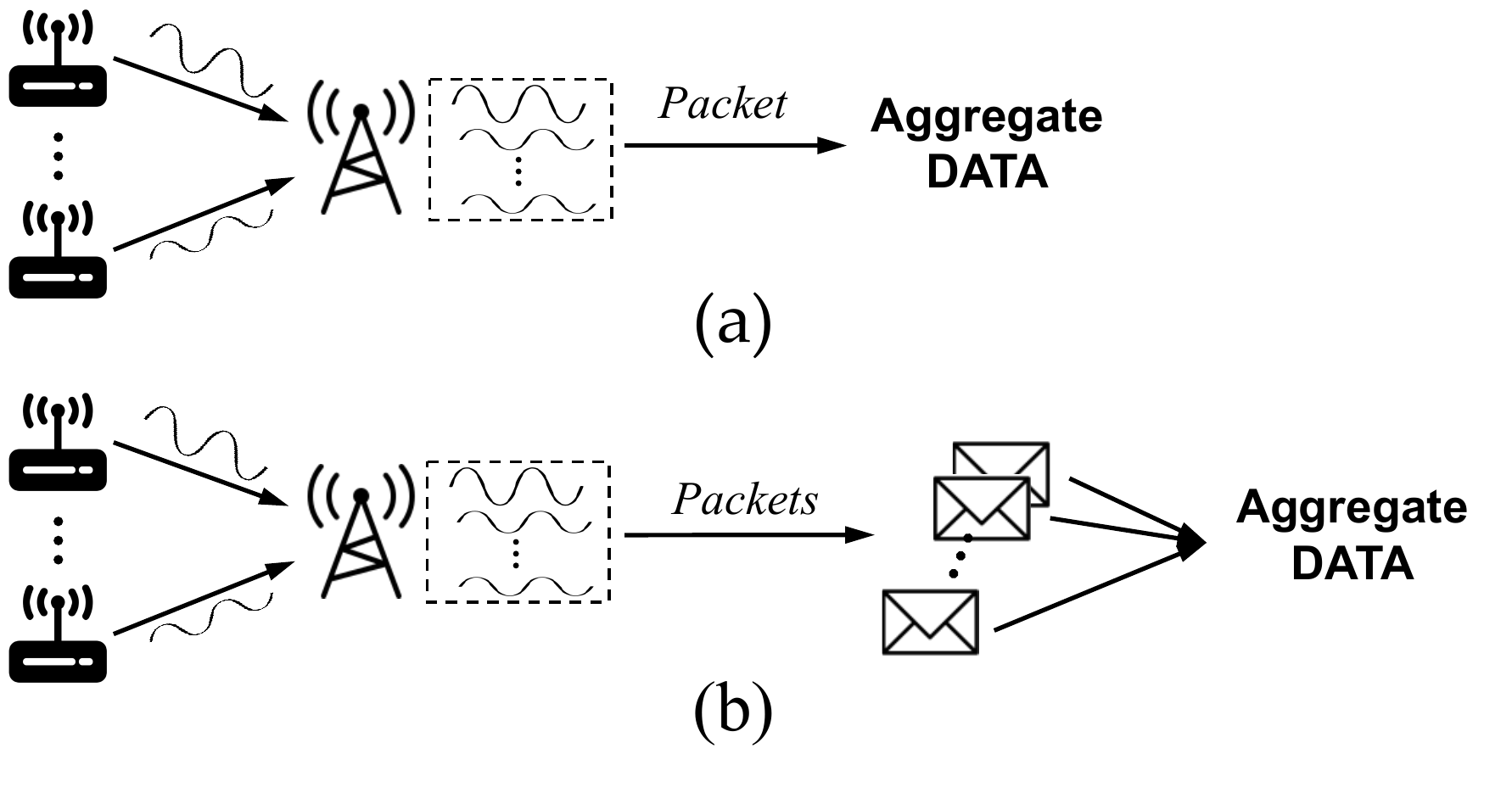}
	\caption{Two physical-layer data aggregation approaches: (a) the analog approach that directly maps the power of the superimposed signal to aggregate data; (b) the digital approach that decodes each user's data, and then aggregates.}
	\label{fig:data_agg}
	\vspace{-0.5cm}
\end{figure}

A possible way to improve the aggregate query speed is to perform \emph{physical-layer data aggregation}. Several LoRa nodes transmit their sensed data simultaneously, and the LoRa gateway computes aggregate data directly from the superimposed signal. This concurrent transmission paradigm allows more transmissions and can improve channel access performance. %
QuAiL \cite{gadre2020quick} is the first work that supports physical-layer data aggregation in LPWAN. Specifically, transmitters modulate their data into the power of the carrier signal, and the gateway directly decodes the aggregate data from the power of the received carrier signal. However, this analog approach as shown in Fig.~\ref{fig:data_agg}(a) is not reliable if transmitters are not phase synchronized and power synchronized. %
Even a small packet time offset (TO) can cause destructive interference, lowering the power of the superimposed signal, which in turn causes large aggregation error. Moreover, this solution must assume that data is not protected by channel codes, since we could not recover aggregate data from aggregated channel-encoded data. Therefore, this solution does not apply to COTS LoRa nodes with Hamming codes.

This paper presents a digital physical-layer data aggregation approach as shown in Fig.~\ref{fig:data_agg}(b) that is compatible with COTS LoRa nodes and only requires modifications in LoRa gateway. LoRa nodes encode their data following the standard channel coding and modulation, and are coordinated to transmit simultaneously for quick aggregation. Moreover, the redesigned gateway decodes the aggregate data from the roughly time-synchronous but phase-asynchronous superimposed signal. 
In particular, the gateway uses a new multi-packet reception (MPR) approach that first decodes each user's data, and then aggregates. 

The physical-layer data aggregation scenario introduces two new problems, where existing MPR methods\cite{wang2019mlora,wang2020oct,shahid2021cic,tongcolora2020,tong2020nscale,xia2019ftrack,hu2020sclora,xu2021pyramid,chen2021aligntrack,eletrebychoir17,xia2021pcube} could not deal with:

\begin{enumerate}
	\item \emph{Small TOs}: Coordinated transmissions are preferable for quick physical-layer data aggregation, and small TOs (less than 10\% of the symbol duration) are possible as demonstrated in \cite{ramirez2019longshot}. %
	However, most MPR methods  \cite{wang2019mlora,wang2020oct,shahid2021cic,tongcolora2020,tong2020nscale,xia2019ftrack,hu2020sclora,xu2021pyramid,chen2021aligntrack} are designed for uncoordinated transmissions where TOs among nodes should be large (at least 20\% of the symbol duration);%
	\item \emph{Co-located Peaks}: LoRa usually performs decoding in the frequency domain, where different data exhibit different peak locations. For concurrent transmissions under small TOs, it is highly possible that their peaks get co-located (more analysis in Section \ref{sec:design:challenges}), making user identification and packet decoding challenging.
	A few MPR methods \cite{eletrebychoir17,xia2021pcube} that may work under small TOs could not handle the co-located peak problem.
\end{enumerate}

\noindent \textbf{Our Approach.} To address these problems, this paper presents a new system called LoRa \underline{P}hysical-layer \underline{D}ata \underline{A}ggregation (LoRaPDA) to realize quick and reliable aggregate queries in LoRa. %
We first present a new approach that is able to combat the co-located peak problem in data symbols under small TOs, and identify peaks (and symbol data) belonging to each node. Our approach is inspired by the observation that the peak value of each node in demodulation windows \emph{almost} remains constant over the whole packet. This is due to the constant channel within the duration of a packet in a static or semi-static environment.
We can first compute the peak values through collided preambles at the beginning of the packet, and then use these values for demodulation in data symbols. In this way, we can reduce the peak identification problem to a sequence decision problem, and address the problem through maximum likelihood (ML) detection. For each symbol, we compute the probabilities of all combined data sequences that match the received signal and choose the sequence that maximizes the probability. %
The approach naturally solves the co-located peak problem, since different users are allowed to have the same data in the generated sequences.

\begin{figure}
	\centering
	\includegraphics[width=0.45\textwidth]{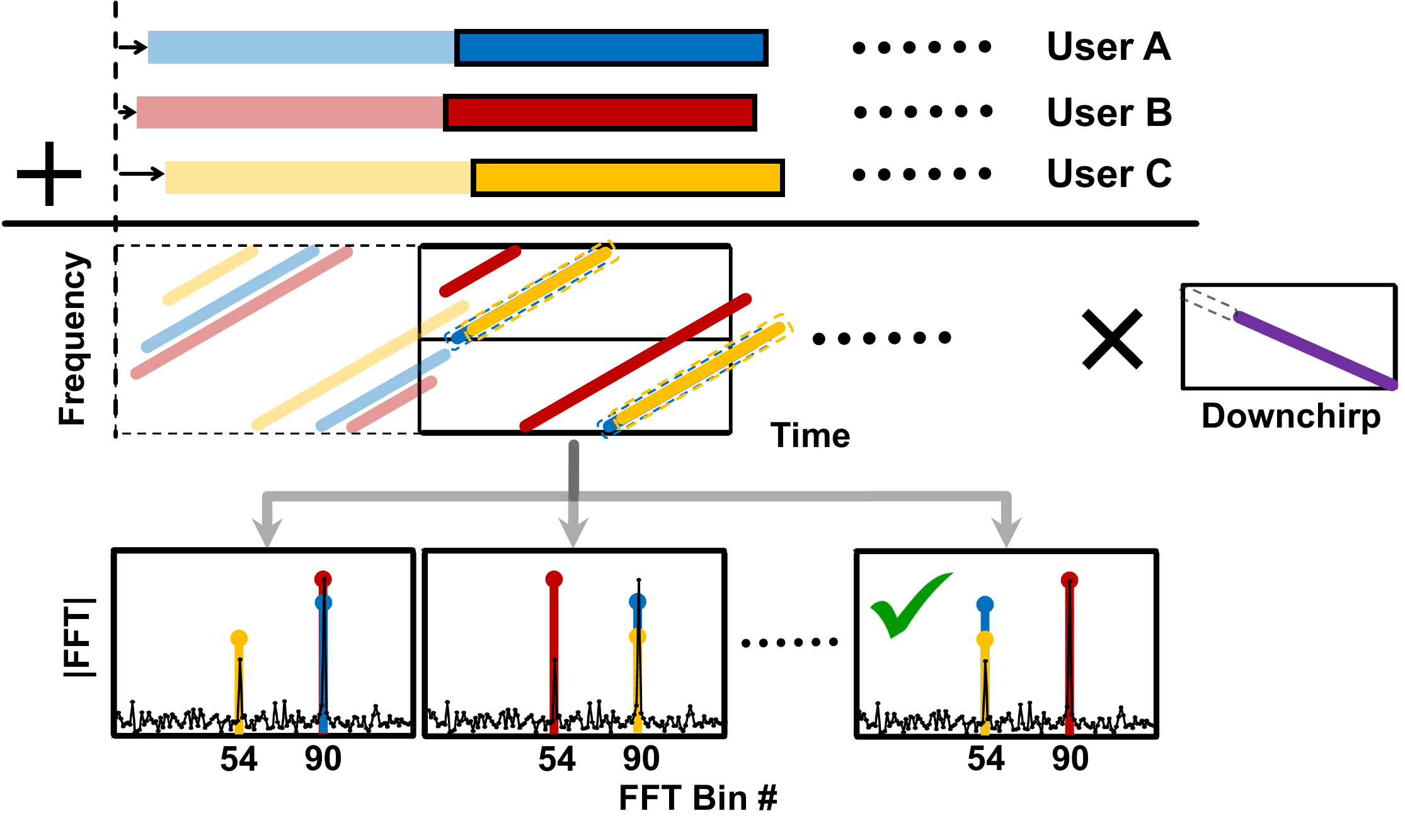}
	\caption{Illustration of LoRaPDA's hard demodulation method. For each symbol, we enumerate all possible data sequences, compute their probabilities and choose the sequence with maximal probability.}
	\label{fig:example}
\end{figure}

The demodulation approach can be illustrated through an example in Fig.~\ref{fig:example}, where three nodes transmit simultaneously with different but small TOs. In the first symbol, there are no collisions since there are three distinct peaks, and it is possible to map peaks to nodes by their amplitude information. In the second symbol, there are only two peaks, and there must be two nodes sharing the same peak. It is hard to associate the co-located peak with the corresponding node by existing MPR methods. For LoRaPDA, we compute the probabilities of all possible sequences, find the last sequence that has the maximal probability, and then map data in the sequence to corresponding nodes. %

Although the idea sounds straightforward, computing the probability that the received peaks match a given transmission sequence is not easy, especially when there are co-located peaks. The peak amplitude of each node remains constant over the whole packet, but the peak phase is not constant due to the existence of CFO and TO. Therefore, co-located peaks have different superimposition results in different symbols. To tackle this challenge, we improve an existing channel and offset estimation algorithm with more fine-grained parameters (i.e., air-channel, CFO, and TO), and present a new signal reconstruction method that accurately tracks the phase impact of CFO and TO to each symbol. %
Simulation results show that the channel estimation mean-squared error is significantly better than the existing algorithm (e.g., 20dB improvement at SNR -5dB$\sim$25dB). 
Trace-driven simulation results show LoRaPDA outperforms other state-of-the-art (SOTA) MPR methods by 5.3$\times$ in terms of physical-layer throughput.%

The above algorithm may work most of the time, but estimation errors and noise may lead to demodulation errors. Taking Fig.~\ref{fig:example} for example, the first sequence is the correct decision, but its probability may be slightly smaller than that of the last sequence. Choosing the maximal probability leads to demodulation errors. 
To combat such errors, LoRaPDA further presents a soft-decision packet decoding algorithm that leverages more demodulation information for channel decoding. In particular, LoRaPDA keeps $K$ sequences with top-$K$ highest probabilities and utilizes a soft-input Hamming decoder for packet decoding. In this way, several sequences that have close probabilities are reserved for packet decoding, and channel codes may be able to correct demodulation errors. 

However, it is not easy to implement such an idea. LoRa adopts the chirp spread spectrum (CSS) modulation where several bits are encoded into a symbol, and the symbol probability is not equivalent to the bit probability that the channel decoder desires. Moreover, LoRa adopts gray mapping for Hamming codes that flips some bits. To tackle these challenges, we develop new mapping algorithms that can convert symbol-level probability to bit-level probability, and get the correct bit probability after gray mapping. Trace-driven simulation results show that the soft decoder (\emph{LoRaPDA-Soft}) can support more users than the hard decoder (\emph{LoRaPDA-Hard}) under the same SNR in practical scenarios. Network-layer trace-driven simulation results show that LoRaPDA-Soft outperforms other SOTA MPR methods by at least 2.1$\times$ using four-user concurrent transmissions at all SNRs.%

Another design is to reduce the high computation complexity of ML detection for real-time decoding. Note that for each symbol, we need to enumerate all possible data sequences, and compute their probabilities. To reduce the number of possible sequences, we leverage the information that the number of peaks is smaller than the number of concurrent users to avoid all enumeration, and present a new enumeration approach with some approximation.
Simulation results show that such approximation achieves almost the same performance as the ML detection at medium to high SNRs. Furthermore, we note that the ML computation for each symbol and the probability computation for each sequence is independent, and thus they can be executed in parallel. Therefore, we implement LoRaPDA on the graphics processing unit (GPU) hardware, and utilize its massive parallel computing capability to accelerate signal processing. Experiment results on the USRP software-defined radio platform show that the decoding time is comparable to the packet duration for two-to-six concurrent transmissions, realizing real-time decoding.

\textbf{Main Contributions.} The main technical contributions of this paper are summarized as follows:
\begin{itemize}
	\item We present an improved channel and offset estimation algorithm and a signal reconstruction method that enables accurate phase tracking in each data symbol.
	\item We present a new symbol demodulation method that uses the ML detection approach to decode concurrent transmissions and combat the co-located peak problem. %
	\item We present a new soft decoder that augments the above demodulation method with more information and improves the channel decoding performance.
	\item We implement and evaluate LoRaPDA on USRP with GPU, demonstrating quick and reliable data aggregation and the possibility of real-time decoding. %
\end{itemize}

The rest of this paper is organized as follows. Section \ref{sec:motivation} introduces the background and motivation. Section \ref{sec:overview} overviews the architecture and workflow of our system. Section \ref{sec:design:hard} and Section \ref{sec:design:soft} present the design details. Experiment results are given in Section \ref{sec:exp}. Section \ref{sec:related} summarizes related works. Section \ref{sec:conclusion} concludes this paper.

\section{Background and Motivation} \label{sec:motivation}

\subsection{LoRa PHY} \label{sec:lora_phy}
LoRa's physical layer adopts the CSS modulation that encodes data bits into chirps with linearly increasing or decreasing frequency, i.e., upchirps or downchirps, respectively. The CSS modulation is robust against interference, low SNRs, and Doppler effects. %

A complete LoRa signal consists of preambles, Start Frame Delimiters (SFDs), and data symbols. Preambles consist of repeated upchirps, and SFDs consist of repeated downchirps. They are used for packet synchronization, channel estimation, and CFO estimation. The frequency of upchirp linearly increases from $-\frac{BW}{2}$ to $\frac{BW}{2}$, occupying the whole bandwidth of $BW \in \{125, 250, 500\}$KHz. A chirp symbol contains $N=2^{SF}$ chips, where $SF \in \{6,7,\dots,12\}$ is called the spreading factor. The upchirp $C$ can be written as $C(t)=e^{j2\pi(\frac{k}{2}t-\frac{BW}{2})t}$, where $k=\frac{BW}{T_s}$ is the gradient of frequency sweeping, and $T_s=\frac{N}{BW}$ is the symbol duration.

LoRa encodes data bits into data symbols by cyclically shifting the upchirp. In particular, it encodes data $s \in \{0,1,\dots,N-1\}$ into signal $x(t)$ with $C(t)$ and an initial frequency shift $f(s)=s\cdot\frac{BW}{N}$, where $x(t)=e^{j2\pi f(s)}C(t)$. CSS enforces the symbol frequency between $-\frac{BW}{2}$ to $\frac{BW}{2}$. Hence the symbol frequency starts from $f(s)-\frac{BW}{2}$, increases linearly with time until it reaches $\frac{BW}{2}$ at $t=t_{fold}$, and folds down to $-\frac{BW}{2}$. Due to the existence of frequency folding, $x(t)$ can be represented as

\vspace{-2.5mm}
\begin{equation}
	\small
	\begin{aligned}
		x(t)&=
		\begin{cases}
			e^{j2\pi(\frac{k}{2}t+f(s)-\frac{BW}{2})t},\ 0\leq t\leq t_{fold}, \\
			e^{j2\pi(\frac{k}{2}t+f(s)-\frac{3BW}{2})t},\ t_{fold}\textless t\leq T_s. 
		\end{cases} \\
		\label{eq:x_t}
	\end{aligned}
	\vspace{-0.2in}
\end{equation}

Let $h$ be the air-channel coefficient, $n$ be the Gaussian noise, and $\delta$ be the CFO between transmitter and receiver. Due to the long symbol duration of the LoRa signal, $h$ is usually treated as a complex scalar. Assume that the demodulation window is perfectly aligned with the transmitter, and the received signal is given by
$y(t) = h\cdot x(t) e^{j2\pi \delta t} +n(t)$.
To demodulate the received signal, a common approach is to first compensate CFO by multiplying $y(t)$ with $e^{-j2\pi \delta t}$, and then dechirp it by multiplying with downchirp $C^*(t)$ %

\begin{equation}
	\begin{aligned}
		y'(t) &= C^*(t)y(t)e^{-j2\pi \delta t} \\
		&= h\left(e^{j2\pi f(s)t}\Pi_1(t) + e^{j2\pi (f(s)-BW)t}\Pi_2(t)\right),
		\label{eq:y_t_d}
	\end{aligned}
\end{equation}
	
\noindent where $u(\cdot)$ denotes the Heaviside step function, $\Pi_1(t)=u(t)-u(t-t_{fold})$ and $\Pi_2(t)=u(t-t_{fold})-u(t-T_s)$ are the rectangular functions. If we apply the Fourier Transform $\mathcal{F}(\cdot)$ to $y'(t)$, two frequency tones will be generated. Under the sampling rate larger than the bandwidth, two tones generate two peaks with frequency $f(s)$ and $f(s)-BW$.	Under the sampling rate equal to bandwidth, these two tones add up at the same frequency $f(s)$, and generate a single peak. %

In addition to the CSS modulation, LoRa adopts Hamming codes to provide error correction. The coding rate $CR \in \{\frac{4}{5},\frac{4}{6},\frac{4}{7},\frac{4}{8}\}$. A complete LoRa transmitter also includes whitening, interleaver, and Gray mapping blocks prior to modulation\cite{tapparel2020open, robyns2018multi}. A complete LoRa receiver includes the corresponding blocks that perform the inverse processing in contrast to the transmitter.

\ignore{
	\begin{figure}[t]
		\setlength{\belowcaptionskip}{0.5cm}
		\centering
		\subfloat[]{%
			\includegraphics[width=0.25\columnwidth]{figures/fig_css_stft1.pdf}
			\vspace{-0.15cm}
			\label{fig:css_stft1}}
		\hspace{-4mm}
		\subfloat[]{%
			\includegraphics[width=0.25\columnwidth]{figures/fig_css_fft1.pdf}
			\vspace{-0.15cm}
			\label{fig:css_fft1}}
		\hspace{-4mm}
		\subfloat[]{%
			\includegraphics[width=0.25\columnwidth]{figures/fig_css_stft2.pdf}
			\vspace{-0.15cm}
			\label{fig:css_stft2}}
		\hspace{-4mm}
		\subfloat[]{%
			\includegraphics[width=0.25\columnwidth]{figures/fig_css_fft2.pdf}
			\vspace{-0.15cm}
			\label{fig:css_fft2}}
		\vspace{-0.2cm}
		\caption{LoRa uses chirp spread spectrum to modulate signals. Above shows the spectrogram of received symbols: (a) (b) base up-chirp, and (c) (d) shifted data chirp symbol.}
		\label{fig:css_example}
		\vspace{-0.75cm}
	\end{figure} 
}

\begin{figure*}[t]
	\setlength{\belowcaptionskip}{0.5cm}
	\centering
	\subfloat[]{%
		\includegraphics[width=0.4\columnwidth]{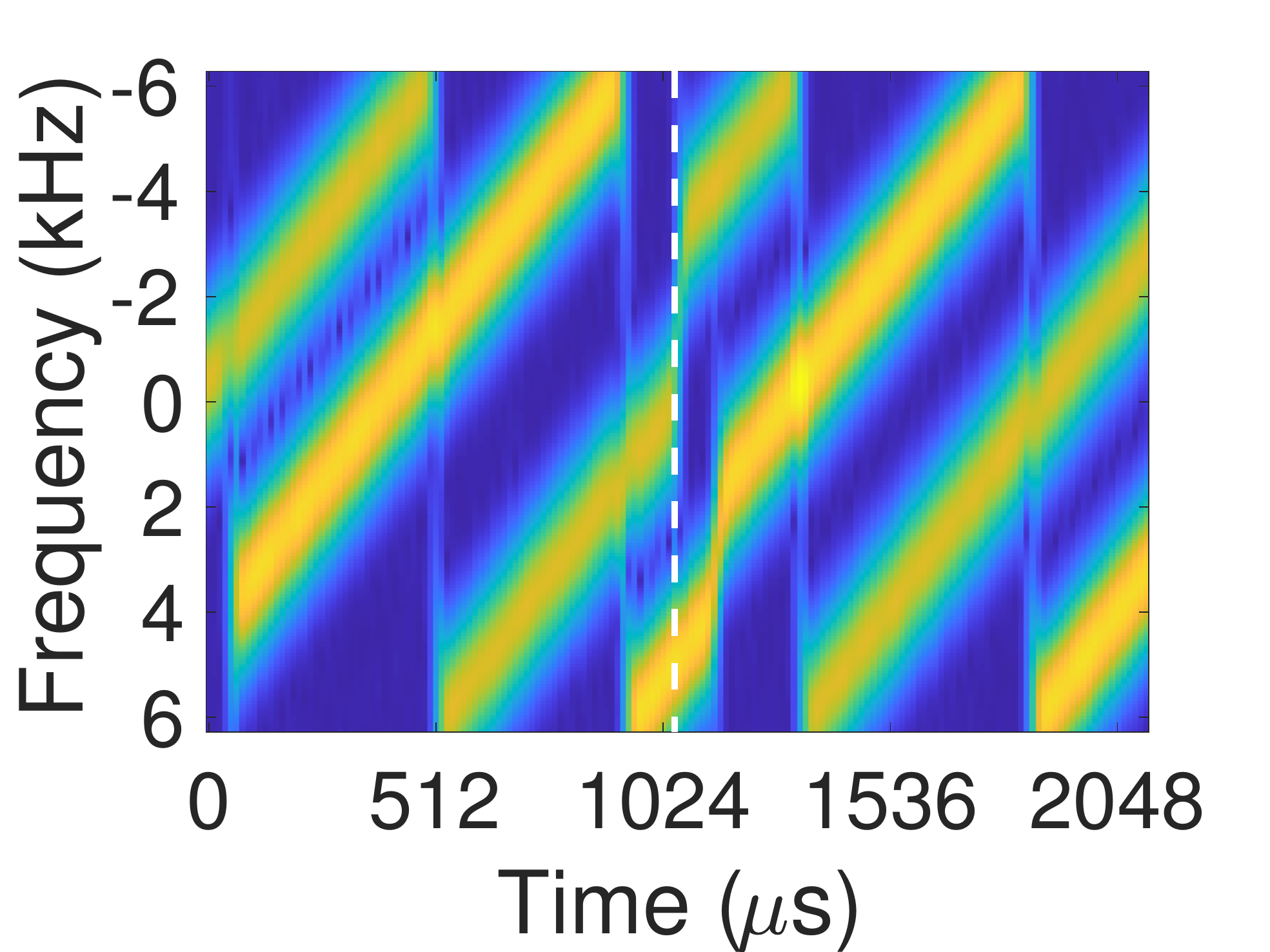}
		\label{fig:2u_stft}}
	\hspace{-3mm}
	\subfloat[]{%
		\includegraphics[width=0.4\columnwidth]{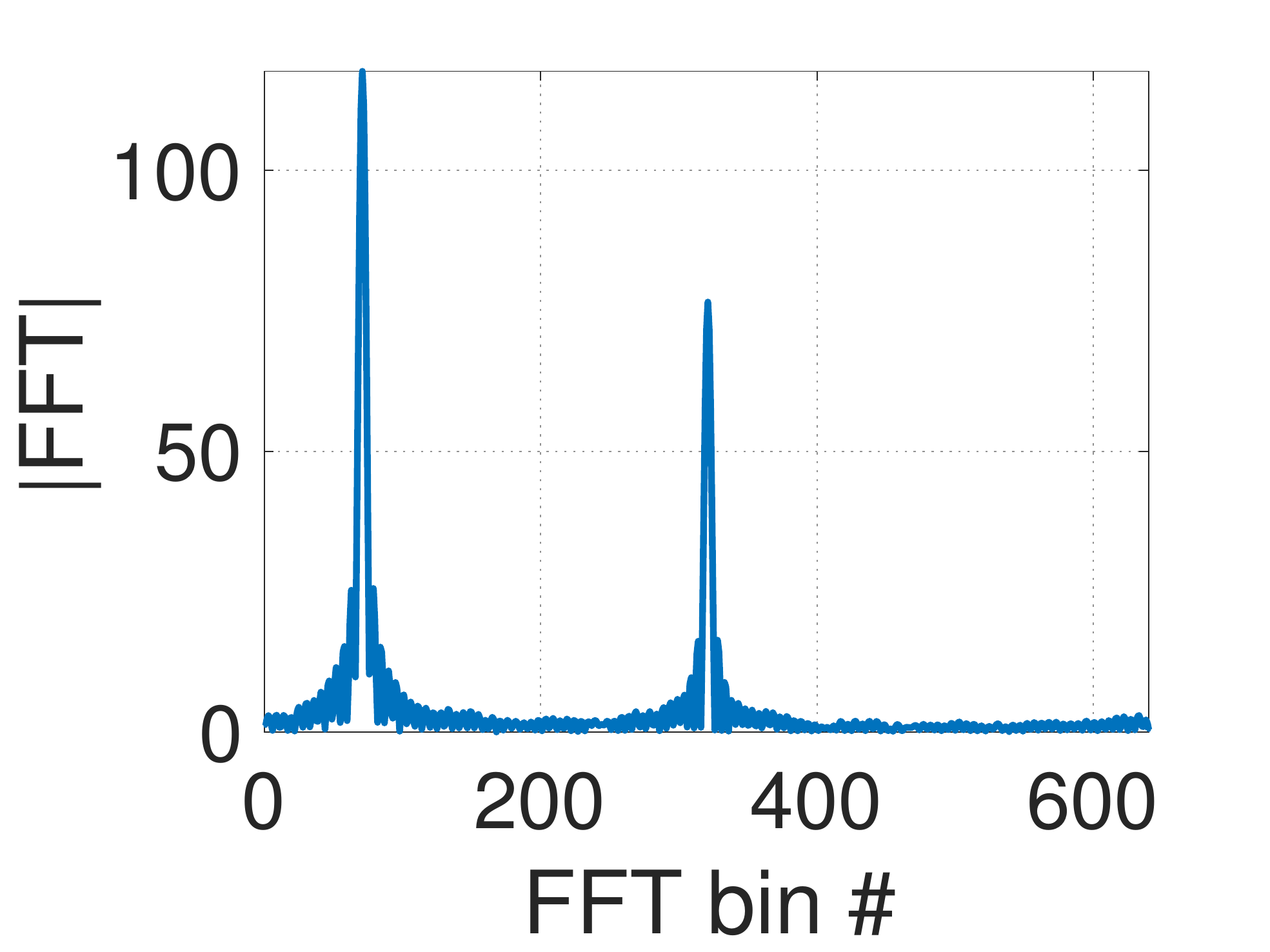}
		\label{fig:2u_fft1}}
	\hspace{-3mm}
	\subfloat[]{%
		\includegraphics[width=0.4\columnwidth]{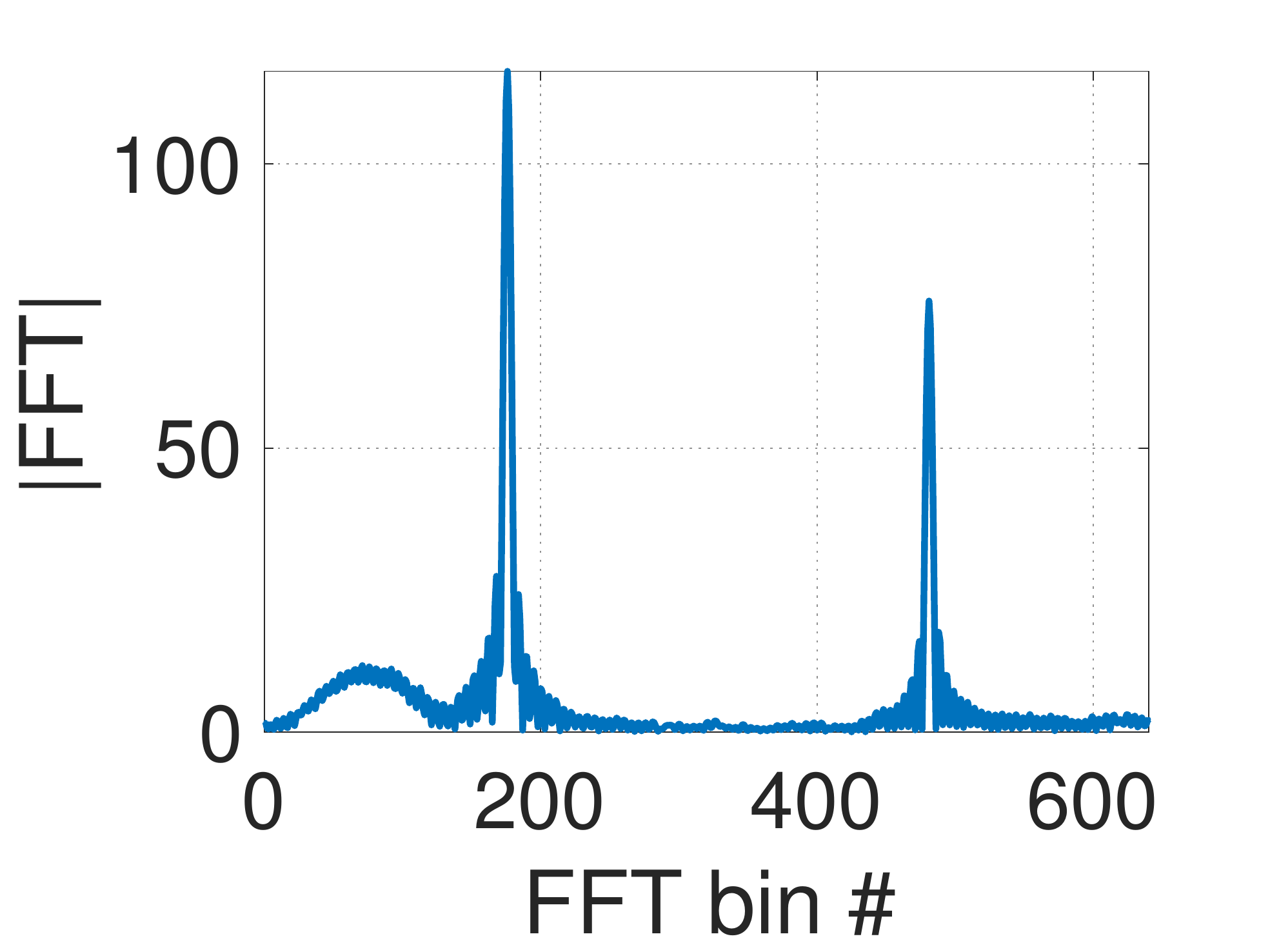}
		\label{fig:2u_fft2}}
	\subfloat[]{%
		\includegraphics[width=0.4\columnwidth]{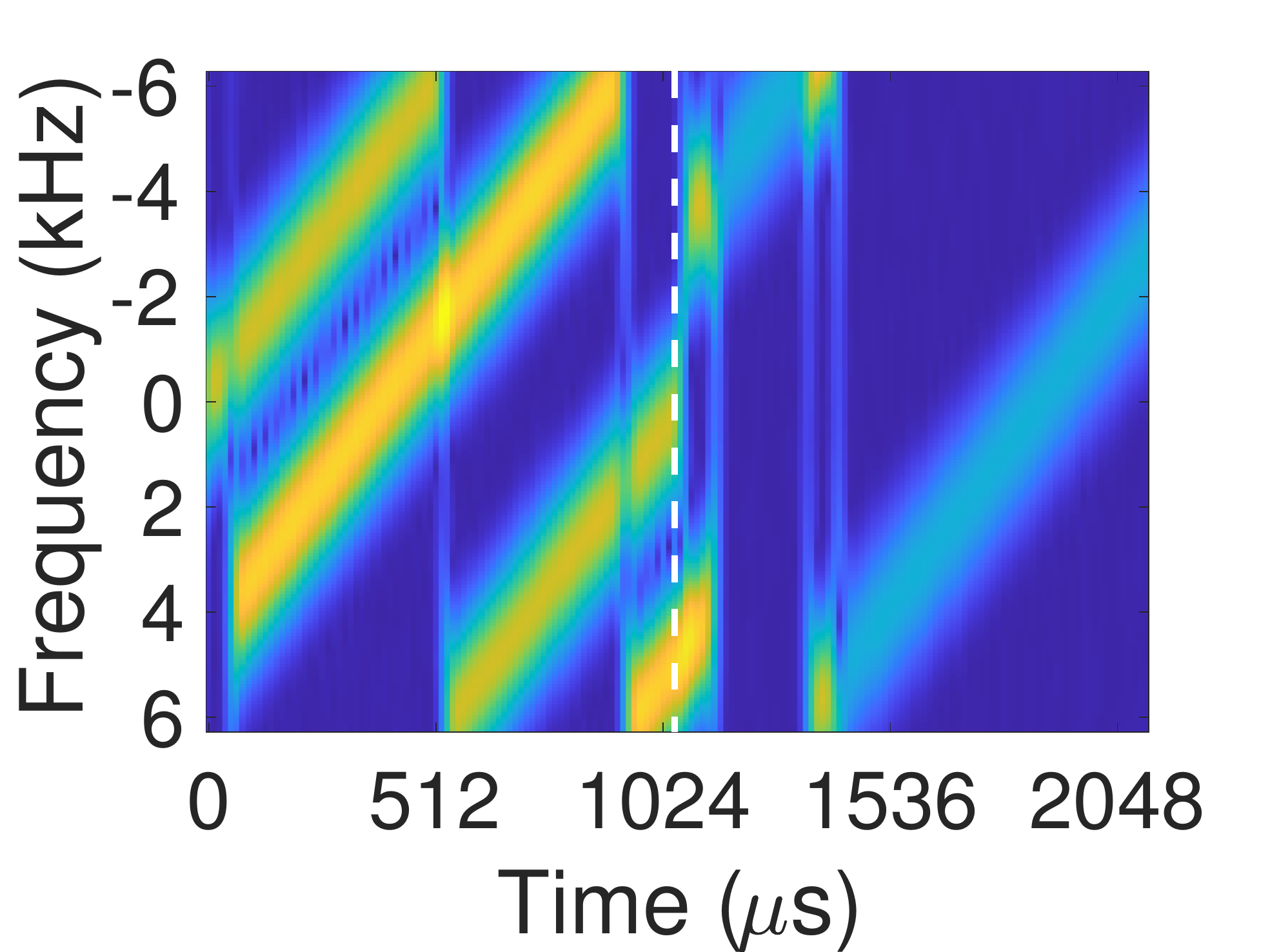}
		\label{fig:sim_colocated_example_stft}}
	\hspace{-3mm}
	\subfloat[]{%
		\includegraphics[width=0.4\columnwidth]{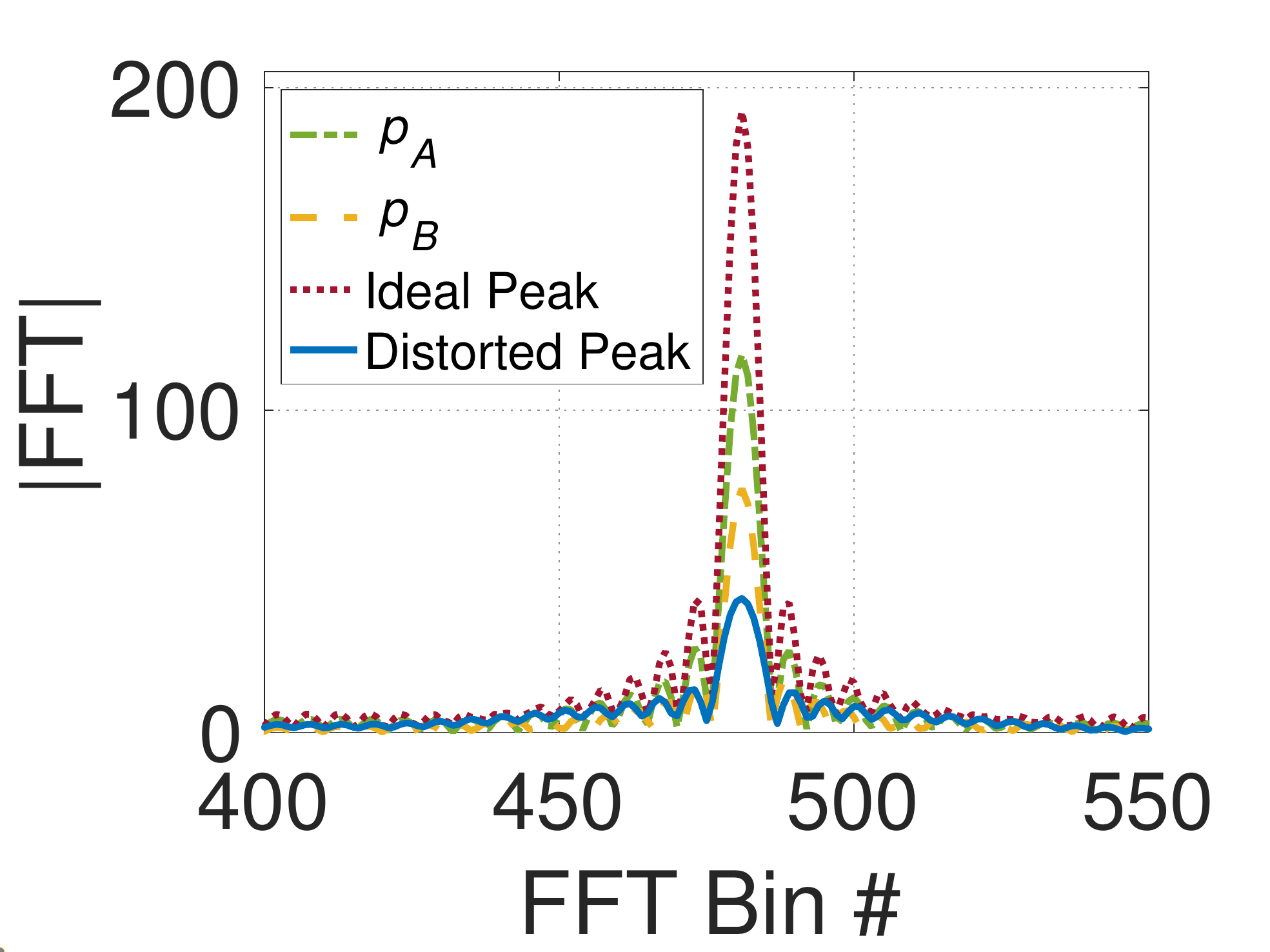}
		\label{fig:fig_sim_colocated_example_fft2}}
	\caption{An example of two-user concurrent transmissions under small TOs: (a) spectrogram of two successive received symbols without co-located peaks; (b) demodulation window of the first symbol; (c) demodulation window of the second symbol; (d) spectrogram of two successive received symbols with co-located peaks in the second symbol; (e) demodulation window of the second symbol where the received superimposed peak is distorted.}
	\label{fig:sim_colocated_example}
\end{figure*} 

\textbf{Concurrent Transmission Model:}
We target the concurrent transmission case that $M$ coarsely synchronized nodes transmit packets simultaneously using the same $SF$. It is possible to realize a synchronization error of less than 2$\mu s$ (less than 1 sample) as described in \cite{ramirez2019longshot} for COTS nodes. Let $x_m$ denote the transmit signal from node $m$, and $y$ denote the collided signal at the gateway. Denote $\delta_m$, $\tau_m$, and $h_m$ as the CFO, TO, and channel coefficients of node $m$, respectively. Assume that the demodulation window is aligned with the first-arrived packet, and thus we have $\tau_m \geq 0$. Each packet contains $N_s$ symbols. Then, the received $i$-th symbol at the gateway can be written as 

\vspace{-4mm}
\begin{equation}
	\small
	\begin{aligned}
		y^{(i)}(t) &= \sum_{m=1}^M h_m x_m^{(i)}(t-\tau_m) e^{j2\pi \delta_mt}+n(t)\\
		&= \sum_{m=1}^M h_m C(t-\tau_m) e^{j2\pi \left(f(s_m^{(i)}) + \delta_m\right)t}+n(t),
	\end{aligned}
	\label{eq:recv_signal}
\end{equation}

\noindent where $s_m^{(i)}$ is the $i$-th transmitted symbol of node $m$, and $i \in \{0,1,\dots,N_s-1\}$. Note that the impacts of CFO and TO on signals are not equivalent. CFO induces a frequency term across all symbol boundaries continuously, whereas TO induces a frequency term whose phase will be reset on the symbol boundaries\cite{Xhonneux2022lorasync}. %

Moreover, due to the linear increasing properties of the CSS modulation, a shift in time of $\tau$ can be interpreted as a shift in frequency of $k\tau$, where $k$ is the gradient of frequency sweeping. Thus, both the CFO and TO can be converted into frequency bin shifts as 

\vspace{-2mm}
\begin{equation*}
	\small
	\begin{aligned}
		\Delta_{\delta_m}=\frac{N\cdot \delta_m}{BW},~\Delta_{\tau_m}=\frac{N\cdot k\tau_m}{BW}=\tau_m \cdot BW.
	\end{aligned}
\end{equation*}

\subsection{Challenges} \label{sec:design:challenges}

The key operation in decoding concurrent transmissions is to match peaks with users (i.e., \emph{user identification}).
However, the physical-layer data aggregation scenario introduces two new challenges, where existing MPR methods\cite{wang2019mlora,wang2020oct,shahid2021cic,tongcolora2020,tong2020nscale,xia2019ftrack,xu2021pyramid,chen2021aligntrack,eletrebychoir17,hu2020sclora,xia2021pcube} that are compatible with COTS LoRa nodes could not deal with:

\textbf{a) User Identification under Small TOs:}
In physical-layer data aggregation, coordinated concurrent transmissions are preferable so that their packet time offset (TO) is small. For example, concurrent transmissions within one sample (i.e., $\sim$2$\mu$s) have been demonstrated in \cite{ramirez2019longshot}.
For $M$ concurrent transmissions, ideally, if we dechirp Eq.~(\ref{eq:recv_signal}) and apply FFT, %
we can obtain $M$ different peaks located at $\mathcal{P}^{(i)}=\{p_{(1)}^{(i)},\cdots,p_{(M)}^{(i)}\}$ with amplitudes $\mathcal{A}^{(i)}=\{a_{(1)}^{(i)},\cdots,a_{(M)}^{(i)}\}$ in the demodulation window, where $p_{(m)}^{(i)} \in [1, 2^{SF}]$. 
The first problem is how to map peaks in $\mathcal{P}^{(i)}$ to corresponding users (nodes). This step is essential for deriving symbol data, because each user is associated with its unique CFO and TO. Assume that there exists an ideal approach, which can rearrange $\mathcal{P}^{(i)}$ to $\mathcal{P}^{(i)}_r=\{p_1^{(i)},\cdots,p_M^{(i)}\}$. Then, we can deduct the bin shifts caused by CFO and TO from the peak locations to get the transmitted symbol data. The $m$-th user's data is given by $s_m^{(i)}=p_m^{(i)}-(\Delta_{\delta_m}-\Delta_{\tau_m})$.

Fig.~\ref{fig:sim_colocated_example}(a)-Fig.~\ref{fig:sim_colocated_example}(c) illustrate the user identification problem under small TOs, where there are two received symbols from two users. In the first symbol and the second symbol, both two peaks exist. We need to group peaks belonging to the same user in these two symbols.

\textbf{b) User Identification under Co-located Peaks:} 
During concurrent transmissions under small TOs, another phenomenon named \emph{co-located peaks} exists. In $\mathcal{P}^{(i)}_r$, some peaks may share the same position (i.e., $p_m^{(i)} = p_n^{(i)}$ for some nodes $m$ and $n$), since nodes randomly choose an initial frequency that represents their data from $N=2^{SF}$ frequencies.
Co-located peaks may add up destructively if their phases are not aligned. %
Fig.~\ref{fig:sim_colocated_example}(d) and Fig.~\ref{fig:sim_colocated_example}(e) present an example of co-located peaks at the second symbol. Ideally, without the impact of the channel, two peaks will add up constructively, resulting in an ideal superimposed peak. In practice, due to the phase misalignment, these two peaks may add up destructively, resulting in a distorted peak.%

Co-located peaks further complicate the user identification problem. Assuming non-co-located peaks, it may be possible to utilize the amplitude information of the channel to map peaks to corresponding nodes. However, co-located peaks invalidate the use of channel amplitude information for user identification, since the amplitude of a co-located peak may be equal to the amplitude of another non-co-located peak, due to the unpredicted phase difference.

\begin{figure}[t]
	\setlength{\belowcaptionskip}{0.5cm}
	\centering
	\subfloat[]{%
        \includegraphics[width=0.5\columnwidth]{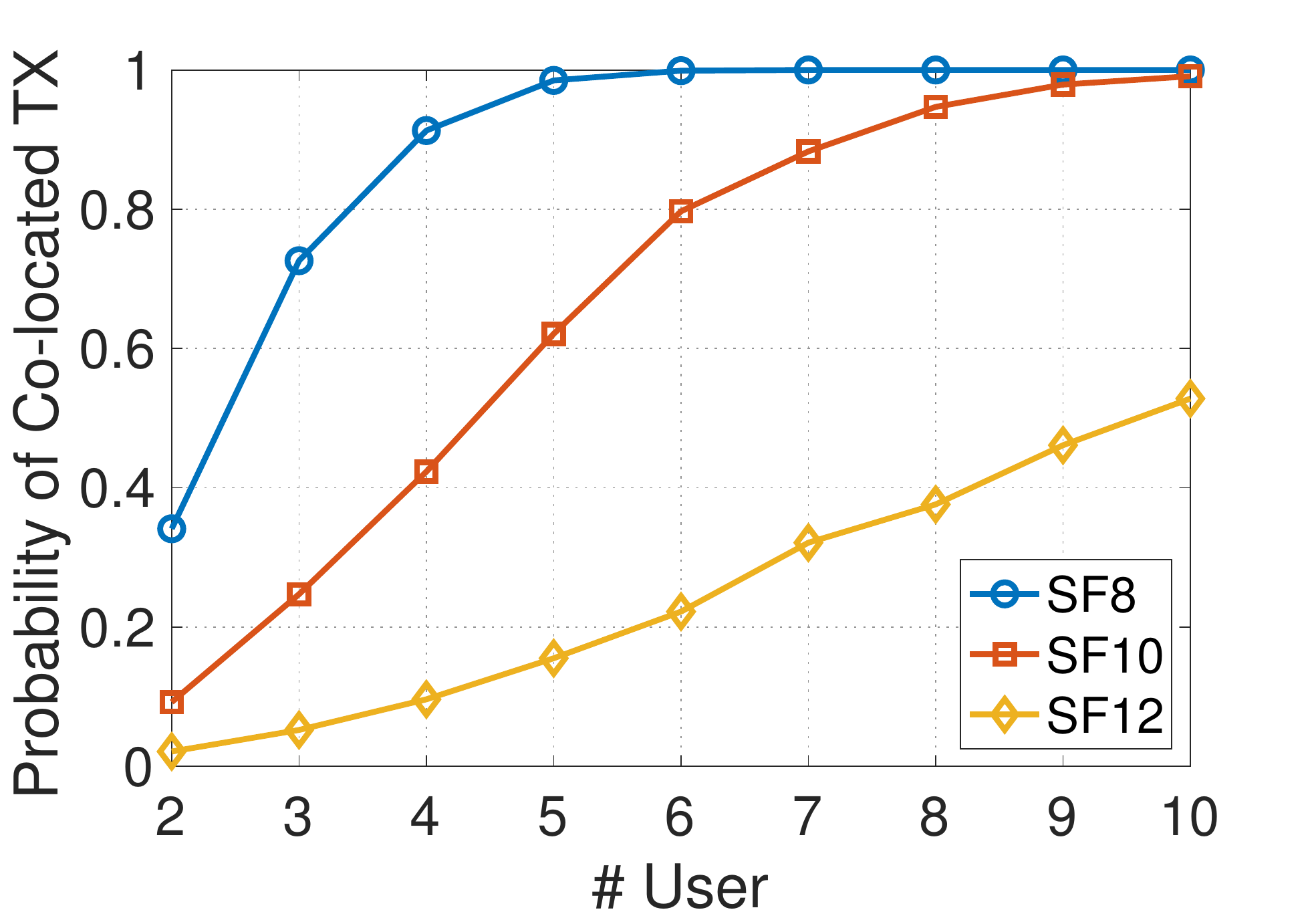}
		\label{fig:sim_pr_colocated}}
	\subfloat[]{%
		\includegraphics[width=0.5\columnwidth]{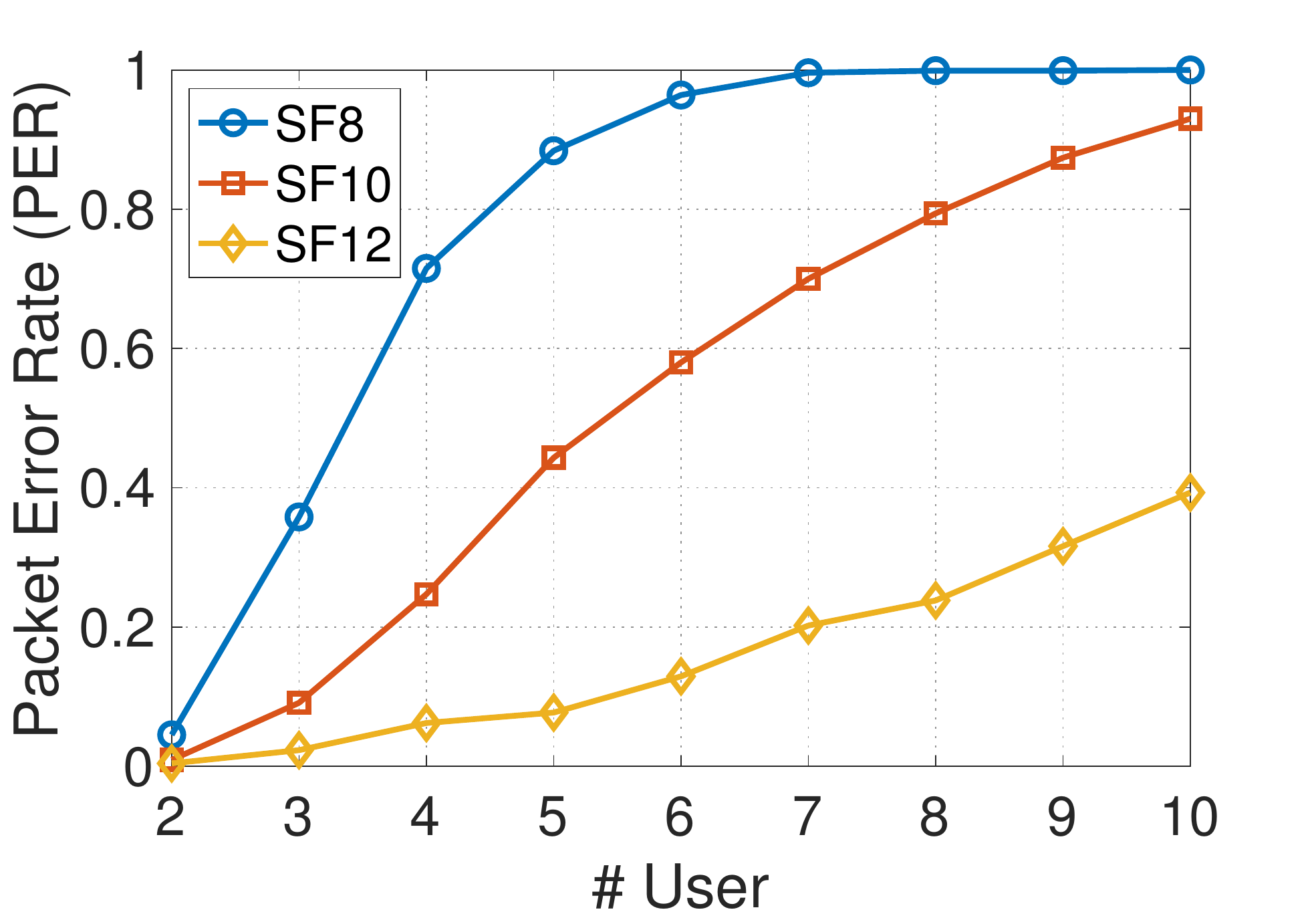}
		\label{fig:sim_colocated_per}}
	\caption{Characterizing the impact of co-located peaks: (a) probability of co-located transmission over the number of users when $SF$=8,10,12; (b) packet error rate using an ideal data mapping method when $SF$=8,10,12.}
	\label{fig:sim_colocated}
\end{figure}

\begin{table*}[t]
	\centering
	\caption{Comparison of LoRaPDA and state-of-the-art MPR methods.  Freq (TO) means that the method uses frequency feature for user identification but the frequency feature is related to TO.}
	\setlength{\tabcolsep}{1.5mm}{
		\begin{tabular}{ccccccc}
			\toprule \textbf{References} & \textbf{Method} & \textbf{Feature} & \textbf{Small TO}  & \textbf{Co-located  Peaks}\\
			\hline
			mLoRa\cite{wang2019mlora} & Time-domain successive interference cancellation & Time & $\times$ & $\times$ \\
			OCT\cite{wang2020oct} & Freq-domain interference cancellation & Freq (TO) & $\times$ & $\times$ \\
			CIC\cite{shahid2021cic} & Freq-domain interference cancellation & Freq (TO) & $\times$ & $\times$ \\
			CoLoRa\cite{tongcolora2020}, NScale\cite{tong2020nscale} & Same peak ratio between two windows & Freq (TO) & $\times$ & $\times$ \\
			FTrack\cite{xia2019ftrack} & Different start position of frequency track & Freq (TO) & $\times$ & $\times$ \\
			SCLoRa\cite{hu2020sclora} & Minimal cumulative spectral coefficient error & Freq (TO) + Power & $\times$ & $\times$ \\
			Pyramid\cite{xu2021pyramid} & Same peak change pattern & Freq (TO) + Power & $\times$ & $\times$ \\
			AlignTrack\cite{chen2021aligntrack} & Largest peak during window sliding & Freq (TO) + Power & $\times$ & $\times$ \\
			Choir\cite{eletrebychoir17} & Same frequency offset & Freq & $\checkmark$ & $\times$ \\
			PCube\cite{xia2021pcube} & Same phase difference of air-channels & Freq + Phase & $\checkmark$ & $\times$ \\
			\hline
			\textbf{LoRaPDA} & Near-ML Demodulation & Freq + Phase + Power & $\checkmark$ & $\checkmark$ \\
			\bottomrule
		\end{tabular}}
		\label{table:vs_mpr}
\end{table*}

\textbf{Impact of Co-located Peaks:}
Furthermore, for concurrent transmissions under small TOs, it is highly possible that their peaks get co-located, making user identification and packet decoding more challenging.
We perform a preliminary study on how often co-located peaks happen in the coordinated transmission scenario. 
Denote a co-located transmission as a packet that contains at least one symbol having co-located peaks. Fig.~\ref{fig:sim_colocated}(a) shows the probability of the co-located transmission with the different number of users who transmit with $BW$=125KHz, $CR=\frac{4}{8}$, and 20-Bytes payload. The $SF$s are set to 8, 10, and 12, leading to 48, 40, and 32 symbols, respectively. The maximal CFO is set to 5KHz, a value measured by \cite{shen2021radio} over many COTS nodes. The maximal TO is set to $10\%\cdot T_s$ (i.e., $\sim$200$\mu$s). Even with bin shifts from CFO and TO, the probability of co-located TX is higher than 80\% when there exist more than three users. Another observation is that the probability of co-located transmission is closely related to $SF$. The probability decreases when $SF$ increases, since the collision domain (i.e., $2^{SF}$) is increased. However, even when $SF$=12, the probability is not negligible.

The next question is whether co-located peaks cause decoding errors after channel decoding. Assume an ideal data mapping algorithm that peak bins in non-co-located symbols can be perfectly mapped to their corresponding users, and peak bins in co-located symbols are mapped to users randomly. Note that in a co-located symbol, a co-located peak may be confused with a non-co-located peak, since the phase misalignment among co-located peaks is uncontrollable. Fig.~\ref{fig:sim_colocated}(b) shows the packet error rate (PER) with different number of users using the algorithm. Co-located symbols have little impact on two-user decoding, since it introduces at most 1 bin error and can easily be corrected by channel codes. The PER is more than 60\% when four users are involved in a transmission, which is unacceptable in practice. Hardly any packets can be correctly decoded when the number of users is larger than five.

\subsection{Limitations of SOTA MPR Methods} \label{sec:mov:sota}

Many MPR methods have been proposed to tackle the collision decoding problem in LoRa as summarized in Tab.~\ref{table:vs_mpr}. Most methods are designed for uncoordinated transmissions, where large TOs are usually assumed (TO $>$ $20\% \cdot T_s$)  to separate and group peaks to users. Moreover, they could not handle co-located peaks.  %

mLoRa\cite{wang2019mlora} leverages distinct TOs of collided signals to extract time-domain interference-free segments, and then performs signal reconstruction and successive interference cancellation (SIC) to resolve collision in correlated segments. 
OCT\cite{wang2020oct} leverages distinct TOs to divide collided signal into segments, and performs peak intersection between consecutive segments (i.e., \emph{frequency-domain interference cancellation}) to extract target data. CIC\cite{shahid2021cic} further extends the idea to more concurrent transmissions. 
However, these approaches rely on interference-free or correlated segments due to distinct TOs that may not exist under small TOs.

CoLoRa\cite{tongcolora2020} leverages the observation that the ratio of two peak values across two successive windows is constant but different for users with distinct TOs for user identification.
NScale\cite{tong2020nscale} further amplifies the effect of distinct TOs by dechirping superimposed signals with non-stationary scaling downchirps and conventional downchirps. However, %
they do not work under small TOs since the ratio would be similar, and the co-located peaks do not lead to constant peak ratios over symbols.

FTrack\cite{xia2019ftrack} slides the demodulation window, and separates packets by detecting the frequency continuity of different users with distinct TOs starting at different positions. 
SCLoRa\cite{hu2020sclora} utilizes the cumulative spectral coefficients that combine both frequency and power features to classify users. %
Pyramid\cite{xu2021pyramid} and AlignTrack\cite{chen2021aligntrack} also utilize the sliding window approach and the different start positions of the peak change pattern for user identification. These sliding-window-based solutions highly rely on the existence of large TOs between packets, which is difficult to be applied under small TOs. Moreover, the co-located peaks would invalidate the same peak change pattern from symbol to symbol.

Some methods may work under small TOs by utilizing frequency domain or phase features to separate symbols, but they are not robust to co-located peaks. Choir\cite{eletrebychoir17} leverages inherently different frequency offset (FO) of each user to identify peaks. That is, peaks with the same fractional frequency bin in the data symbol probably belong to the same packet that has the same fractional FO calculated via preambles. Even though this approach can work without distinct TOs, it also suffers from the co-located peak problem where fractional FO computation is no longer accurate. PCube\cite{xia2021pcube} leverages peak phase differences on multi-antennas among nodes for user identification and packet decoding.  However, the phase differences are not reliable when two peaks share the same frequency bin.

\section{LoRaPDA Overview} \label{sec:overview}

\ignore{
	{\bf Challenges.} It introduces three key challenges in aggregate queries: 
	
	\begin{itemize}
		\item How to estimate accurate phase rotation under the impact of CFO and TO? 
		\item How to correctly reconstruct symbols as the phase changes across symbols?
		\item How to improve the decoding performance under the residual error of symbol recovery? 
	\end{itemize}
}

Given the above limitation of the SOTA MPR approaches, we present LoRaPDA with a new MPR approach that can work under both small TOs and co-located peaks.
Fig.~\ref{fig:example} illustrates LoRaPDA's key operations. The key idea is to perform maximum likelihood (ML) detection for each symbol. In particular, LoRaPDA reconstructs signals corresponding to all possible data sequences for each symbol, and calculates their probabilities based on the received signal. To support ML detection, channel and offset estimation and symbol demodulation are essential. To improve detection performance, soft channel decoder is used.

\begin{figure}
	\centering
	\includegraphics[width=0.48\textwidth]{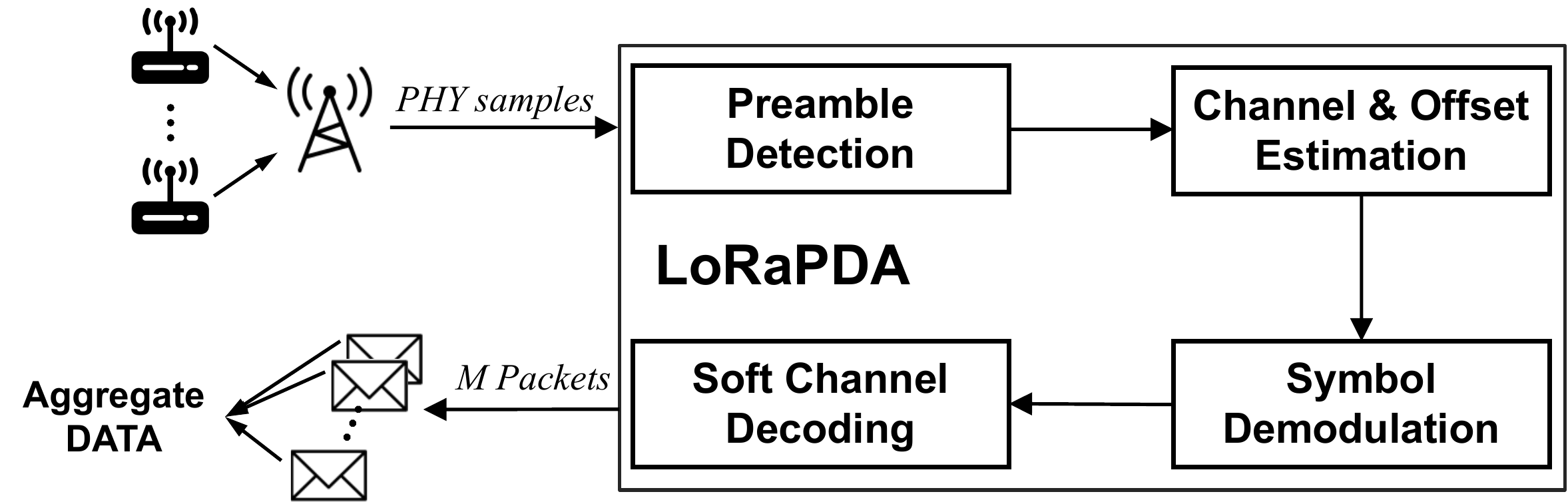}
	\caption{Building blocks of LoRaPDA, and the workflow to perform physical-layer data aggregation.}
	\label{fig:arch}
\end{figure}

Fig.~\ref{fig:arch} shows all building blocks and the workflow of LoRaPDA. Assume that there are $M$ LoRa nodes and one LoRa gateway. The gateway is specially designed for the data aggregation application, while LoRa nodes are COTS nodes. LoRaPDA operates in the following steps: 

\begin{enumerate}
	\item {\emph{Preamble detection.}} We adopt the common approach to detect preambles: we correlate received samples with base upchirp and align our demodulation window with the first-arrived packet. Then we move the demodulation window by a symbol duration each time.
	
	\item {\emph{Channel and offset estimation (\textsection\ref{sec:design:offset_and_channel_estimation}).}} For each user, we estimate CFO and TO using preambles and SFDs, reconstruct the preamble signals with accurate phases, and then perform least-square to estimate air-channel. %
	
	\item {\emph{Symbol demodulation (\textsection\ref{sec:design:symbol_recovery}).}} After extracting valid peaks, we can then reconstruct signals corresponding to each user data sequence for each symbol, compute the probability with given received signals, and find $K$ sequences with the top-$K$ largest probabilities.
	
	\item {\emph{Soft channel decoding (\textsection\ref{sec:design:soft}).}} To achieve higher performance, we use soft-decision decoding with probabilistic demodulation outputs and a soft-input channel decoder for Hamming codes.
	
	\item {\emph{Data aggregation.}} After decoding data of each node from the superimposed signal, we perform data aggregation directly and save the aggregate result only. 
\end{enumerate}

\ignore{
	\subsection{Challenges} \label{sec:design:challenges}
	\begin{itemize}
		\item How to align the demodulation window with the first-arrived packet?
		\item How to recover all symbols in each window as different peaks will interfere or even co-locate with each other?
		\item How to decode packets and meanwhile be compatible with LoRa encoder?
	\end{itemize}
	
	\begin{figure}
		\centering
		\includegraphics[width=160pt]{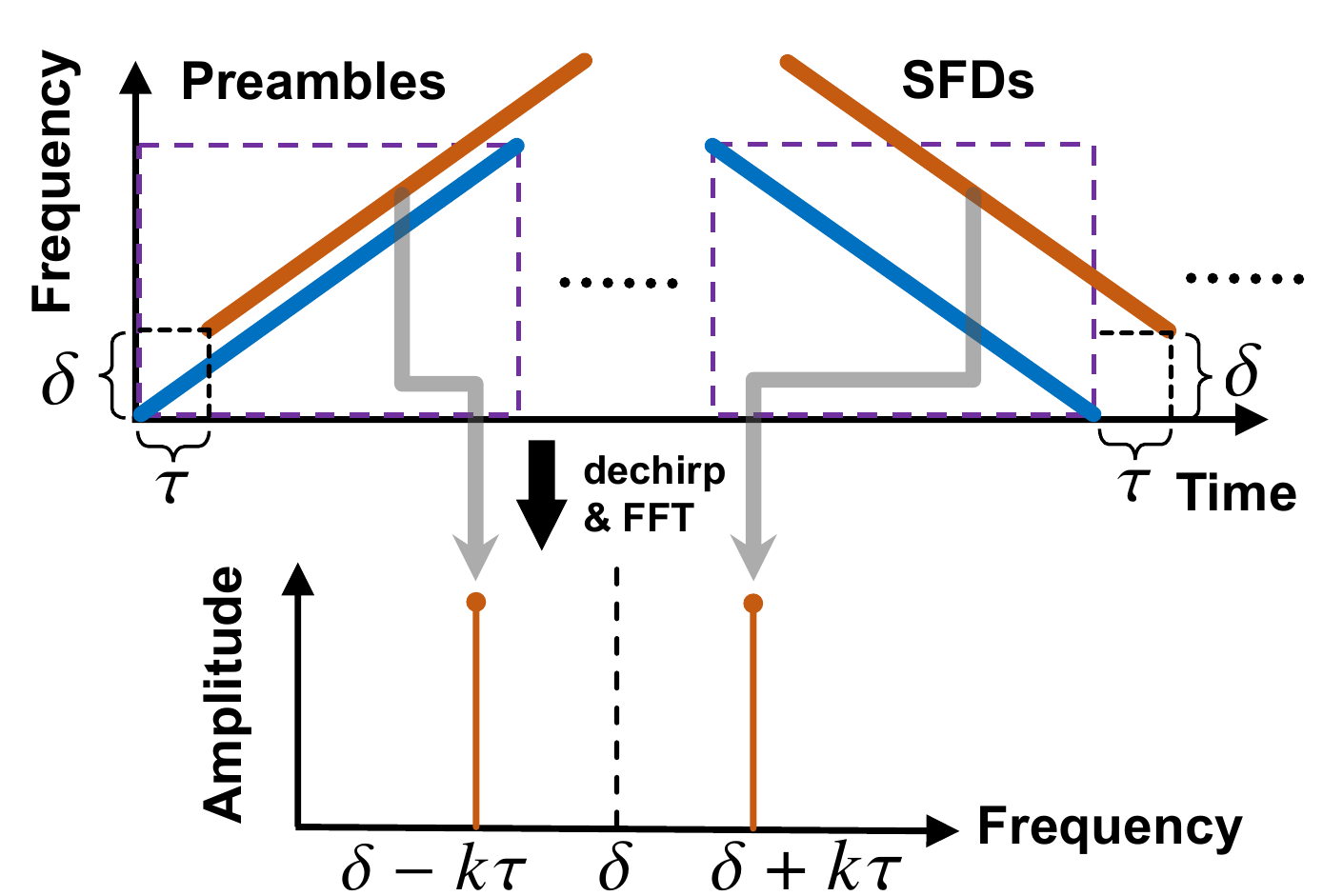}
		\caption{Coarse CFO and TO estimation with preambles and SFDs.}
		\label{fig:coarse_offset_est}
		\vspace{-0.5cm}
	\end{figure}
	
	\subsection{Preamble Detection} \label{sec:design:detection}
	
	\emph{1) Chirp Correlation:}
	
	\emph{2) Window Alignment:}
}

\section{LoRaPDA Hard Decoder} \label{sec:design:hard}
In this section, we first present the channel and offset estimation method, and then present the symbol hard demodulation method.

\subsection{Channel and Offset Estimation} \label{sec:design:offset_and_channel_estimation}
To precisely reconstruct the signal that corresponds to a user data sequence, we need to estimate the phase offsets caused by TO $\tau$, air-channel $h$, and the phase rotation caused by CFO $\delta$ precisely. Our key idea is to first estimate CFO and TO using preambles and SFDs as in \cite{tongcolora2020, chen2021aligntrack}, and then improve the least-square estimation method proposed by \cite{eletrebychoir17} to estimate air-channel. Different from \cite{eletrebychoir17}, we use a more accurate model that captures the phase offsets induced by CFO and TO, and perform the least-square estimation in the frequency domain which is more resistant to noise.

Take two-user concurrent transmissions as an example:
\begin{equation}
	y(t) = h_1 x_1 (t-\tau_1) e^{j2\pi \delta_1 t} + h_2 x_2 (t-\tau_2) e^{j2\pi \delta_2 t} + n(t).
	\label{eq:y_t}
\end{equation}
We first get $\tilde{\delta}_1$, $\tilde{\delta}_2$, $\tilde{\tau}_1$ and $\tilde{\tau}_2$, reconstruct $x_1 (t-\tau_1) e^{j2\pi \delta_1 t}$ and $x_2 (t-\tau_2) e^{j2\pi \delta_2 t}$, and then get $\tilde{h}_1$ and $\tilde{h}_2$. %

\emph{1) CFO and TO Estimation:} We utilize both the preambles and SFDs to estimate CFO and TO of each user. Assume that TO between the start of the demodulation window and the signal is $\tau$. %
CFO causes the same phase shift $\delta$ to preambles and SFDs. In contrast, TO causes negative frequency shifts $-k\tau$ in preambles and positive frequency shifts $k\tau$ in SFDs. For preambles, it results in a peak at frequency $f^{u} = \delta-k\tau$. For SFDs, it results in a peak at frequency $f^{d} = \delta+k\tau$. Thus, CFO and TO can be calculated by $\tilde{\delta} = \frac{f^u+f^d}{2}$ and $\tilde{\tau} = \frac{f^d-f^u}{2k}$, respectively. 

Since user $1$ and user $2$ have different TO/CFO, we can have two peaks $f^u_1, f^u_2$  in preambles, and two peaks $f^d_1, f^d_2$ in SFDs. We use their amplitude information to classify user $1$ and user $2$, and then compute $\tilde{\delta}_1$, $\tilde{\delta}_2$, $\tilde{\tau}_1$ and $\tilde{\tau}_2$.

\emph{2) Preamble Signal Reconstruction:} With estimated CFOs and TOs, we can compute air-channels that best fit Eq.~(\ref{eq:y_t}). Given that preambles are base upchirps with symbol data $s=0$, we can reconstruct preambles of user $1$ and user $2$ by $$E = \left[ \mathcal{R}(0,\tilde{\delta}_1,\tilde{\tau}_1,1) \ \ \mathcal{R}(0,\tilde{\delta}_2,\tilde{\tau}_2,1) \right],$$ 
where $\mathcal{R}(s, \delta, \tau, h)$ is the symbol-level signal reconstruction function defined below. The signal reconstruction function is used in both channel/offset estimation and symbol demodulation.

The reconstructed signal can be written as

	\begin{equation}
		\mathcal{R}(s, \delta, \tau, h)=r_\tau\left(hC(t)e^{j2\pi (f(s)+\delta) t}\right),
		\label{eq:r}
	\end{equation}
	
\noindent where $r_\tau$ denotes the right zero-filling operation with $\tau$ samples shift, modeling the effect of TO. Note that due to the phase-continuous features of CFO, phase rotation caused by $\delta$ can be easily added corresponding to the sample index inside the symbol.
	
The most difficult part in Eq.~\ref{eq:r} is that $\tau$ may not be integer to be zero-filled, since the transmitted time-domain signals are continuous. To address this problem, we use the upsampling and downsampling technique. That is, we first modulate a chirp signal $C_{\gamma}(t)e^{j2\pi f(s)t}$ with an oversampling rate $OSR=\gamma$. Then we add CFO $\delta$ and air-channel coefficient $h$ to the signal, and right shift the signal for $\lfloor\gamma\tau\rfloor$ samples with zero-filling at the left. That is, the TO of $\tau$ is amplified to $\lfloor\gamma\tau\rfloor$ in the signal. Finally, we downsample the signal with a downsampling rate $DSR=\gamma$ to obtain the reconstructed time-domain signal. In our implementation, we find that the oversampling rate $OSR=10$ is a good trade-off in terms of performance and computation overhead, and thus we use it for signal reconstruction.

Since the signal reconstruct function is invoked intensively in LoRaPDA, it is time-consuming to compute Eq.~\ref{eq:r} each time. To reduce computation overhead, we pre-modulate chirp signals with $s\in[0, N-1]$ and $OSR=\gamma$, and store them locally to avoid redundant computation in our implementation.

\emph{3) Air-Channel Estimation:} With $y(t)$ and $E$, we can apply the least-square estimation to get $h_1$ and $h_2$. We perform the least-square estimation in the frequency domain. The closed-form formula can be represented as $$H = \left[\tilde{h}_1 \ \tilde{h}_2 \right] = (E_f^TE_f)^{-1}E_f^T\mathcal{F}(y),$$ 
where $E_f=\mathcal{F}(E\cdot C^*)$ is the FFT results of dechirped reconstructed preambles.%

\begin{figure}[t]
	\setlength{\belowcaptionskip}{0.5cm}
	\centering
	\subfloat[]{%
		\includegraphics[width=0.5\columnwidth]{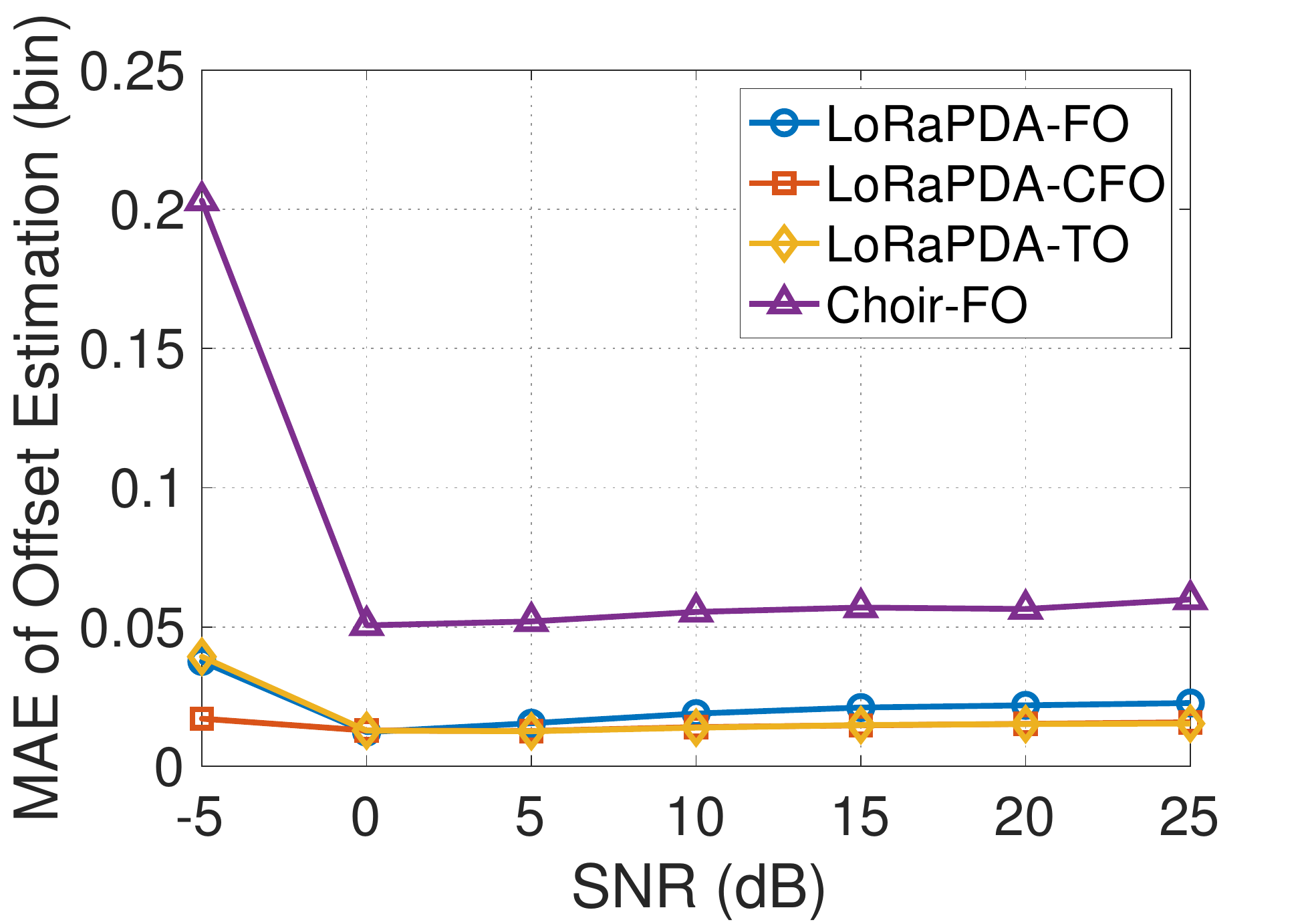}
		\label{fig:sim_pream_est_mae}}
	\subfloat[]{%
		\includegraphics[width=0.5\columnwidth]{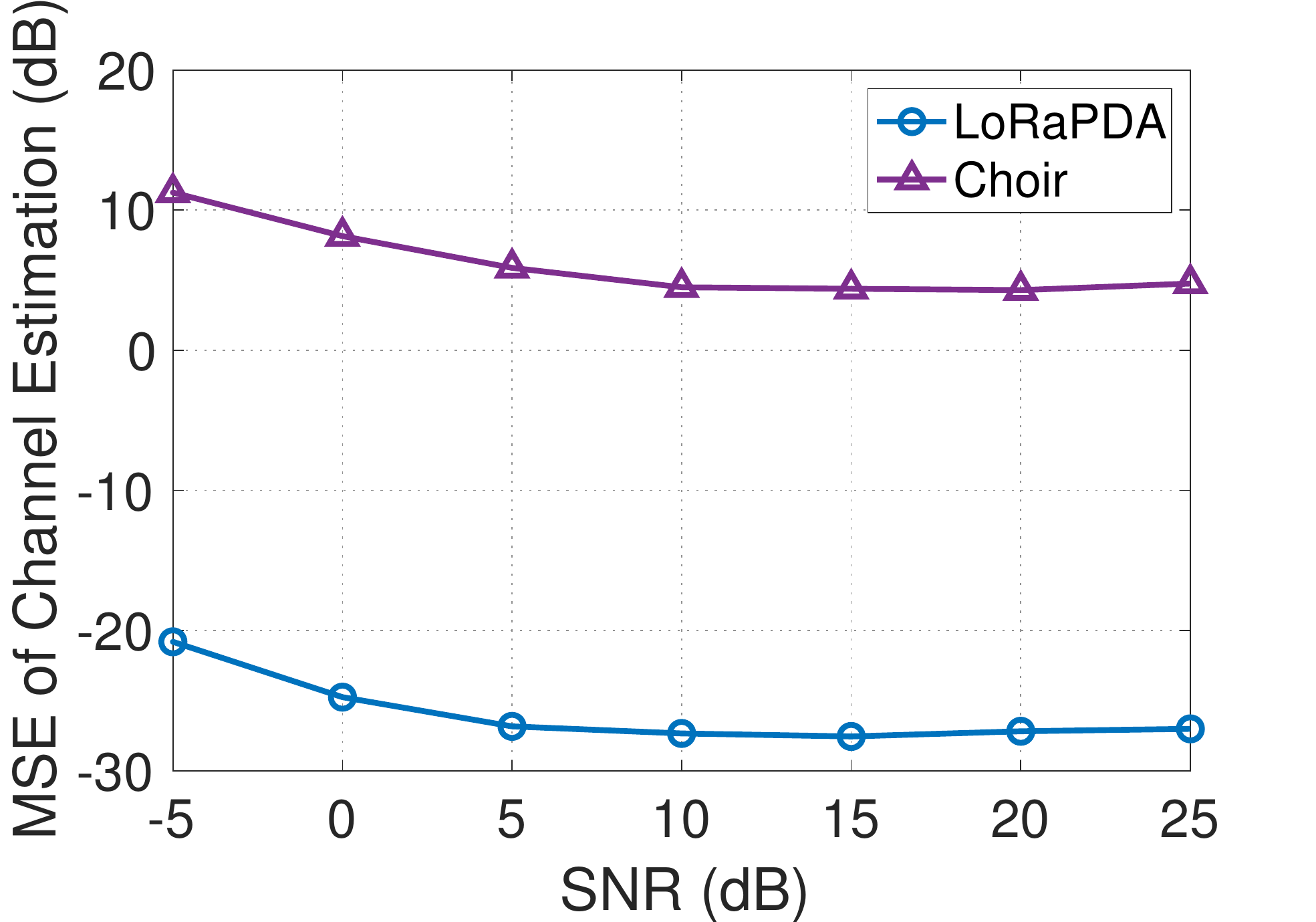}
		\label{fig:sim_pream_est_mse}}
	\caption{Channel and offset estimation results with two-user concurrent transmissions: (a) Mean absolute error (MAE) of offset estimation; (b) Mean-squared error (MSE) of air-channel estimation.}
	\label{fig:sim_pream_est}
\end{figure}

\ignore{
\emph{2) Fine Estimation:} Till now, we have coarsely estimated the air-channel and the offset introduced by CFO/TO. To precisely reconstruct the symbols of each user, we need to estimate CFO, TO and air-channel coefficients as accurately as possible. Intuitively, we can measure the difference between the received symbols and the reconstructed symbols based on coarsely estimation results in the frequency domain as $J(\tilde{\delta}_1,\tilde{\tau}_1; \tilde{\delta}_2,\tilde{\tau}_2) = \left( \left| \mathcal{F}(y \cdot C^*) \right| - \left| H\cdot E_f \right| \right)^2$.

The residual function can be minimized if we find the CFO/TO that best fits the observed signal $\mathcal{F}(y \cdot C^*)$. We can then compute and update the offset coefficients via
\vspace{-0.2cm}

\begin{equation}
	(\delta_m,\tau_m) = \arg\mathop{\min}_{\delta_m\in(\tilde{\delta}_m-\epsilon,\tilde{\delta}_m+\epsilon) \atop \tau_m \in (\tilde{\tau}_{m}-\epsilon,\tilde{\tau}_{m}+\epsilon)} J(\delta_m,\tau_m),
	\label{eq:optimization}
\end{equation}

\noindent where $m \in \{1,2\}$, and $\epsilon$ is the FFT bin size. Exhaustively searching all the time offsets and frequency offsets can be computationally expensive. In our implementation, we apply stochastic gradient-descent algorithms on Eq.~(\ref{eq:optimization}) with several randomly chosen initial points that are likely to converge to the global minimum.
} %

\textbf{Comparison with Choir\cite{eletrebychoir17}:}
Although Choir can also estimate offset and air-channel coefficients using preambles, there are two limitations in estimating air-channel coefficient. First, Choir aggregates TO and CFO as frequency offset (FO) in the model, and uses FO to represent the signal phase. However, CFO and TO have different impacts on the signal phase as discussed in Section \ref{sec:motivation}. The inaccurate phase modeling may have little impact on the FO estimation since the shift still exists even though their superimposition is not accurate, but can have a strong impact on the air-channel coefficient estimation given that the least-square estimation is applied to the reconstructed signal. Instead, LoRaPDA models and estimates CFO and TO separately, and tries to reconstruct the signal with accurate phases. Second, Choir gets FO directly from the peak position. The FO estimation is not accurate, and may degrade the air-channel estimation performance in the least-square estimation. Instead, LoRaPDA leverages preambles and SFDs for CFO and TO estimation, which is more accurate as shown in \cite{tongcolora2020, chen2021aligntrack}.

We use simulations to investigate their performance. 
The $SF$ and $BW$ are set to 10 and 125KHz, respectively. The maximal CFO and TO are 5KHz and 10\%$\cdot T_s$, respectively. We generate 1000 packets with random CFO and TO. Fig.~\ref{fig:sim_pream_est} shows the channel and offset estimation results of two-user concurrent transmissions compared with Choir. 
As shown in Fig.~\ref{fig:sim_pream_est_mae}, 
The MAE of FO derived by Choir remains larger than 0.05 bins even at high SNR, and increases to 0.2 bins when SNR is -5dB, while LoRaPDA achieves higher accuracy, since the MAE of both the CFO and TO is below 0.025 bins across all SNRs. 
Fig.~\ref{fig:sim_pream_est_mse} shows that the channel estimation mean squared error (MSE) of LoRaPDA is at least 20 dB better than that of Choir. The large improvement is due to LoRaPDA's accurate modeling of CFO and TO for air-channel estimation. Note that Choir uses non-coherent detection for packet decoding, and its unsatisfactory channel estimation performance does not impact packet decoding. Instead, LoRaPDA uses coherent detection, and thus requires higher channel estimation accuracy.

\ignore{
	\begin{table}[ht]
		\footnotesize
		\centering
		\caption{Offset and channel estimation comparison.}
		\begin{tabular}{cccccc}
			\toprule
			Method & Domain & CFO & TO & Channel Estimation \\
			\hline
			Choir\cite{eletrebychoir17} & Time & $\checkmark$ & $-$ & TDLS \\
			\textbf{LoRaPDA} & Frequency & $\checkmark$ & $\checkmark$ & FDLS \\
			\bottomrule
		\end{tabular}
		\label{table:vs_choir}
		\vspace{-0.4cm}
	\end{table}
}

\ignore{
	\begin{figure}[t]
		\setlength{\belowcaptionskip}{0.5cm}
		\vspace{-0.3cm}
		\centering
		\subfloat[]{%
			\includegraphics[width=0.45\columnwidth]{figures/fig_sim_peak_frac_to_example_distorted.pdf}
			\vspace{-0.15cm}
			\label{fig:sim_peak_frac_to_example_distorted}}
		\quad
		\subfloat[]{%
			\includegraphics[width=0.45\columnwidth]{figures/fig_sim_peak_frac_to_example_mag_sum.pdf}
			\vspace{-0.15cm}
			\label{fig:sim_peak_frac_to_example_mag_sum}}
		\caption{Peak Distortion Example: distorted peak occurs when the chirp and demodulation window are misaligned (assume $\tau=0.25$), and we address it with FFT magnitude summation.}
		\label{fig:fig_sim_peak_frac_to_example}
		\vspace{-0.8cm}
	\end{figure}
}

\subsection{Symbol Demodulation} \label{sec:design:symbol_recovery}
Symbol demodulation is the key component in LoRaPDA. It includes the following steps:

\emph{1) Peak Extraction:} To recover symbol data belonging to each user, we need to first extract valid peaks in each demodulation window. As we align the demodulation window with the first-arrived user and move the window symbol by symbol, different TOs result in inter-symbol interference, which may lead to erroneous peak extraction.

\ignore{
	\begin{itemize}
		\item First, as we align the demodulation window with the first-arrived packet and move it window by window, different TOs of other packets result in inter-symbol interference. This may lead to misjudgment of extractions.
		\item Second, packets with TOs result in distorted peaks, impacting the extraction of accurate locations.
		\item Third, due to spectrum leakages of sinc function resulting from applying FFT to the rectangular window function, the side lobes around peaks may be misidentified as peaks.
	\end{itemize}
}

To avoid the inter-symbol interferences from the previous symbols, we adopt truncated downchirps $C^*_{tr}$ to dechirp signals. As shown in Fig.~\ref{fig:example}, we cut a small piece of downchirp (in purple) to multiply with symbols under small TO. In our implementation, the truncated length is set according to the measured time synchronization accuracy. 

After FFT, the demodulation window has several peaks. In the worst case, each bin can be a possible location of a peak, since the amplitude of a co-located peak can be low under adverse phase differences. Computing all $N^M$ combinations is expensive. To reduce complexity, we set a threshold to filter peaks, and denote the locations (indexes) of peaks in the $i$-th window as $\{p_1^{(i)}, p_2^{(i)}, \cdots, p_V^{(i)}\}$, 
where $V$ is the number of valid peaks, and $V < N$ with high probability. The threshold can be measured and set beforehand using noise signal.

Another problem is that TO $\tau$ may affect the accuracy of peak extraction. If the packet is accurately aligned with the demodulation window ($\tau=0$), two frequency tones in Eq.~(\ref{eq:y_t_d}) (i.e., $f(s)$ and $f(s) - BW$) will add up constructively, resulting in a single peak. When the demodulation window is misaligned with $\tau>0$, two tones will rotate by different phases after Fourier Transform, i.e., $e^{j2\pi f(s)\tau}$, and $e^{j2\pi (f(s)-BW)\tau}$. Then two peaks with different phases will add up destructively, resulting in peak distortions \cite{tong2020nscale}. %
To overcome this problem, we oversample the signal to expand the frequency range, and only add the magnitude of two pairing tones $f$ and $f-BW$, which results in a distinct peak even when $\tau$ exists. %

\emph{2) Likelihood Mapping:} The collided symbols have several peaks in each demodulation window. To recover the exact data of collided symbols, our approach is to find the data sequence that best fits the observed frequency peaks. 

Let $\mathbf{\Omega}[i]=\{\mathbf{A}^1[i],\mathbf{A}^2[i],\cdots,\mathbf{A}^W[i]\}$ be all possible sequences in the $i$-th symbol, where $\mathbf{A}^{\omega}[i]=(a_1^{\omega}[i],\cdots,a_M^{\omega}[i])$ is a data sequence of all nodes. 
Each node has at most $V$ possible values in a window, i.e., $a_m^{\omega}[i]\in \{p_1^{(i)}, \cdots, p_V^{(i)}\}$, and $W = V^M$. Note that due to existence of CFO and TO, the symbol data of the $m$-th node is given by $a_m^{\omega}[i]-(\Delta_{\delta_m}-\Delta_{\tau_m})$.

For the $i$-th demodulation window, the FFT result in the $i$-th window can be denoted by $Y[i]=\mathcal{F}\left(y^{(i)}(t)\cdot C^*_{tr}\right)$. Our goal is to compute likelihood probabilities $\mathcal{L}[i]=\{L_{\mathbf{A}^{1}}[i], \cdots, L_{\mathbf{A}^{W}}[i]\}$ which indicate how well the sequence fits the observed signal $Y[i]$ in the frequency domain. 

For the sequence $\mathbf{A}^{\omega}[i] \in \mathbf{\Omega}[i]$ in the $i$-th symbol, the reconstructed collided signal can be represented as $$\sum_{m=1}^M \mathcal{R}(a_m^{\omega}[i]-(\Delta_{\delta_m}-\Delta_{\tau_m}),\delta_m,\tau_m,h_m),$$
where signal reconstruction function $\mathcal{R}()$ is defined in Section \ref{sec:design:offset_and_channel_estimation}. 
Denote $\tilde{Y}_{\mathbf{A}^{\omega}}[i]$ as the FFT result of the reconstructed signal from $\mathbf{A}^{\omega}[i]$. Therefore, under the zero-mean additive Gaussian noise $n\sim \mathcal{N}(0,\sigma^2)$, we can write the probability in a frequency bin $j$ as

\vspace{-4mm}
\begin{equation}
	\small
	Pr(Y^j[i] \mid \tilde{Y}^j_{\mathbf{A}^{\omega}}[i]) = \frac{1}{\sqrt{2\pi}\sigma}exp\{ -\frac{\left|Y^j[i] - \tilde{Y}^j_{\mathbf{A}^{\omega}}[i]\right|^2}{2\sigma^2}\}.
	\label{eq:pr_bin_j}
\end{equation}

\noindent Since all frequency bins are independent, the log-likelihood probability of $\mathbf{A}^{\omega}[i]$ is given by

\vspace{-7mm}
\begin{equation}
	\small
	\begin{aligned}
		L_{\mathbf{A}^{\omega}}[i] 
		&= log \prod_{j \in [0,N_{\mathcal{F}}-1]} \frac{1}{\sqrt{2\pi}\sigma}exp\{ -\frac{\left|Y^j[i] - \tilde{Y}^j_{\mathbf{A}^{\omega}}[i]\right|^2}{2\sigma^2}\} \\
		&= log\frac{1}{\sqrt{2\pi}\sigma}exp\{ -\frac{\sum\limits_{j \in [0,N_{\mathcal{F}}-1]} \left|Y^j[i] - \tilde{Y}^j_{\mathbf{A}^{\omega}}[i]\right|^2}{2\sigma^2}\} \\
		&\propto - \sum\limits_{j \in [0,N_{\mathcal{F}}-1]} \left|Y^j[i] - \tilde{Y}^j_{\mathbf{A}^{\omega}}[i]\right|^2,%
	\end{aligned}
\label{eq:log_pr_a}
\end{equation}

\noindent where $N_{\mathcal{F}}$ is the number of FFT bins.

\emph{3) Demodulation: } The simplest demodulation algorithm is to perform hard demodulation. Given $\mathcal{L}[i]$, we can find the sequence $\mathbf{A}^{\mathds{1}}[i]=(a_1^{\mathds{1}}[i],\cdots,a_M^{\mathds{1}}[i])$ in $\mathbf{\Omega}[i]$ with the maximum likelihood probability (equivalent to the minimal distance), where $\mathds{1}$ denotes the index with the maximal probability. Then, we can repeat the above steps for all symbols, and get $\{\mathbf{A}^{\mathds{1}}[0], \cdots, \mathbf{A}^{\mathds{1}}[N_s-1]\}$. Note that $\mathbf{A}^{\mathds{1}}[i]$ contains data for all $M$ users. By extracting each user's data from all symbol sequences, we can construct the packet bits for each user, and use the traditional channel decoder to get source bits.

\todo{
}

\begin{figure}[t]
	\setlength{\belowcaptionskip}{0.5cm}
	\centering
	\includegraphics[width=0.75\columnwidth]{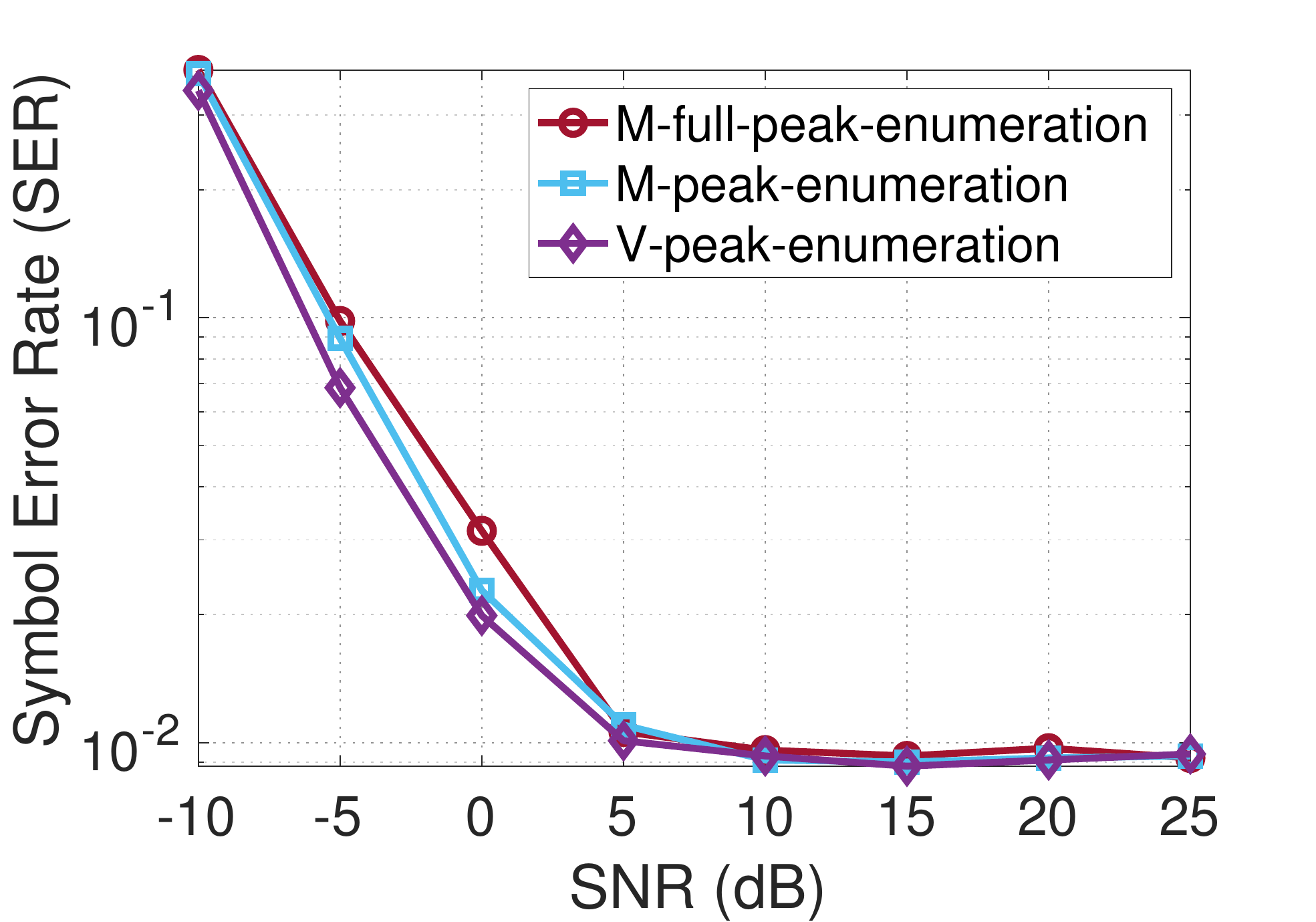}
	\caption{Demodulation performance of different enumeration algorithms under four-user concurrent transmissions with known air-channel, CFOs and TOs.}
	\label{fig:sim_diff_enum_algo}
\end{figure}

\textbf{Optimization:} Exhaustively searching all sequences (i.e., $V^M$) in the threshold-based enumeration method (i.e., \emph{$V$-peak-enumeration}) can be computationally expensive. Note that $M$ users can have at most $M$ peaks. Thus, we have a new enumeration algorithm (i.e., \emph{$M$-peak-enumeration}) that considers $M^M$ sequences, reducing the computation.

We can make further approximations to reduce the computation complexity. Specially, we set the number of possible peaks from $1$ to $M$, and select them from the descending order of the bins' amplitude. Furthermore, we enforce that all selected peaks must be used. The algorithm can be called \emph{$M$-full-peak-enumeration}. Then, the number of combinations is given by $\sum_{V=1}^M f(M,V)$, where $f(m,v)=f(m-1,v) \cdot v+f(m-1,v-1) \cdot v$, $f(\cdot,1)=1$ and $f(v,v)=v!$. The number of combinations is greatly reduced. Taking six (four)-user concurrent transmissions for example, the number of combinations is reduced from 46656 (256) to 4683 (75), nine (three) times lesser.

We perform simulations to compare the demodulation performance of the three enumeration methods using symbol error rate (SER) as the metric. We assume four-user concurrent transmissions, and assume an ideal channel and offset estimation. The $SF$ and $BW$ are set to 10 and 125KHz, respectively. The maximal CFO and TO are 5KHz and 10\% $\cdot$ $T_s$, respectively. 
As shown in Fig.~\ref{fig:sim_diff_enum_algo}, the \emph{$M$-full-peak-enumeration} algorithm achieves almost the same performance as the threshold-based \emph{$V$-peak-enumeration} algorithm at SNR$\geq$5dB, and has a small gap at SNR$\in$[-10,5]dB. Given the considerably reduced complexity, we adopt the \emph{$M$-full-peak-enumeration} algorithm in our implementation.

Another issue is that the computation complexity may still be too high for real-time implementation. To address this issue, we note that the computation of each sequence and the computation of each symbol are dependent, thus we can use the advanced graphics processing unit (GPU) hardware with massively parallel computing capability to accelerate the signal processing (GPU has been used in accelerating 5G baseband signal processing \cite{nvidia5gsdk}). In particular, we use each thread on GPU to compute the probability of one sequence. In this way, we manage to realize a real-time LoRaPDA receiver. The details are presented in Section~\ref{sec:exp:real-time}.

\begin{algorithm}[tb!] 
\caption{Symbol-to-Bit Probability Conversion}
\begin{algorithmic}[1]          %
\Require                    %
Top-$K$ sequence
$\mathcal{A}_m^K[i]=\{a_m^{\mathds{1}}[i],\cdots,a_m^{\mathds{K}}[i]\}$ with corresponding probabilities $\mathcal{L}^K[i]=\{L_{\mathbf{A}^{\mathds{1}}}[i],\cdots,L_{\mathbf{A}^{\mathds{K}}}[i]\}$;
\Ensure                		%
Probabilities of $SF$ bits that is equal to zero $P_{s_m^{[i]}}=$ $\{P_{d_{SF-1}},\cdots,P_{d_0}\}$;

\For {$n = [0, SF)$}
    \State $idx_0=\{bitget(\mathcal{A}_m^K[i],n)==0\}$; \Comment{$n$-th bit is 0}
    \State $idx_1=\{bitget(\mathcal{A}_m^K[i],n)==1\}$;
    \State $l_0=sum(\mathcal{L}^K[i][idx_0])$; \Comment{sum the log likelihood}
    \State $l_1=sum(\mathcal{L}^K[i][idx_1])$;
    \State $P_{d_n}=\frac{l_1}{l_0+l_1}$; \Comment{normalization}
\EndFor
\end{algorithmic}
\label{algo:per_bit_pr}
\end{algorithm}

\ignore{
	\begin{figure}
		\centering
		\includegraphics[width=250pt]{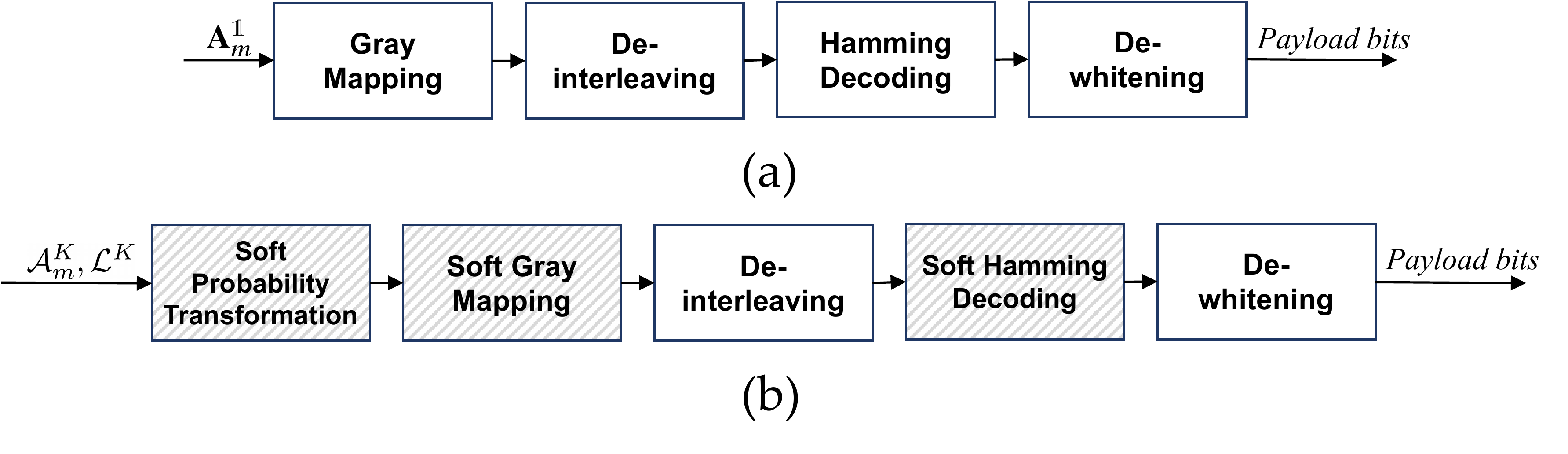}
		\caption{Comparisons of (a) standard LoRa decoder, and (b) soft-decision decoder of LoRaPDA.}
		\label{fig:soft_arch}
		\vspace{-0.5cm}
	\end{figure}
	
	\begin{figure}
		\centering
		\includegraphics[width=80pt]{figures/fig_soft_gray_minus_1.pdf}
		\caption{An example of -1 shift probability transformation. The probability that $d_2$ is zero after -1 shift is equal to two cases (1) $d_2=1$, $d_1=d_0=0$, and (2) $d_2=0$, $d_1\&d_0\neq 0$.}
		\label{fig:soft_gray_minus_1}
	\end{figure}
}

\section{LoRaPDA Soft Decoder} \label{sec:design:soft}

In this section, we first overview the LoRaPDA soft decoder, including background, motivation and challenges for adopting a soft decoder, and then present the details.

\subsection{Overview} 

The above hard-decision decoding approach may lose some information when selecting the sequence with maximal probability. The sequence with maximal probability $\mathbf{A}^{\mathds{1}}[i]$ may not be the optimal sequence $\mathbf{A}^{opt}[i]$, since the noise is inevitable (e.g., channel estimation). 
It may be better to keep some sequences whose probabilities are close-to-maximal and leverage the channel codes to decode the best codeword with probability inputs. Such a method is called \emph{soft-decision decoding}, and is often used in Convolutional codes, Turbo codes, and LDPC codes, which can provide better performance than hard decoder. 

In this way, we can keep a sequence set $\mathcal{A}^K[i]=\{\mathbf{A}^{\mathds{1}}[i],\cdots,\mathbf{A}^{\mathds{K}}[i]\}$ with top-$K$ likelihood probabilities $\mathcal{L}^K[i]=\{L_{\mathbf{A}^{\mathds{1}}}[i],\cdots,L_{\mathbf{A}^{\mathds{K}}}[i]\}$ for the $i$-th symbol. By extracting different nodes from $\mathcal{A}^K[i]$, we can get the top-$K$ sequence set for node $m$ as $\mathcal{A}_m^K[i]=\{a_m^{\mathds{1}}[i],\cdots,a_m^{\mathds{K}}[i]\}$ with  probabilities $\mathcal{L}^K[i]$. Then we can combine the sequence set of all symbols as $\{\mathcal{A}_m^K[0],\cdots,\mathcal{A}_m^K[N_s-1]\}$ with corresponding probabilities $\{\mathcal{L}^K[0],\cdots,\mathcal{L}^K[N_s-1]\}$.

Unfortunately, the standard LoRa channel decoder does not support soft-decision channel decoding. 
For a LoRa receiver, in addition to the CSS demodulation, the receiver also performs Gray mapping, deinterleaving, Hamming decoding, and dewhitening that map symbols to data bits. Gray mapping result is given by $s_g=s'\oplus(s'>>1)$, where $s'=(s-1)\%N$ is the $-1$ shift of $s$. LoRa uses a deinterleaver to distribute (up to $SF$) bit errors over multiple symbols. %
Codeword length is controlled by coding rate $CR \in \{\frac{4}{5},\frac{4}{6},\frac{4}{7},\frac{4}{8}\}$. LoRa adopts $(n_c, k)$ Hamming codes to provide error detection and correction, where $k=4$ is the data bit length. The $(6,4)$ and $(8,4)$ Hamming codes are the punctured version and the extended version of the standard $(7,4)$ Hamming code, respectively. LoRa uses even parity-check code for $CR=\frac{4}{5}$. Finally, payload bits are obtained by XOR-ing Hamming decoding outputs with a pseudo-random sequence at dewhitening.

The soft channel decoder needs to address three new problems:
a) symbol demodulator outputs symbol-level probabilities, where a symbol consists of $SF$ bits, while soft channel decoder needs bit-level probabilities, b) even if we can get bit-level probabilities, gray mapping that changes the bit sequence value would make the input invalid, and c) a soft Hamming channel decoder is required.

\begin{figure} [t]
    \setlength{\belowcaptionskip}{0.5cm}
    \centering
    \includegraphics[width=0.5\textwidth]{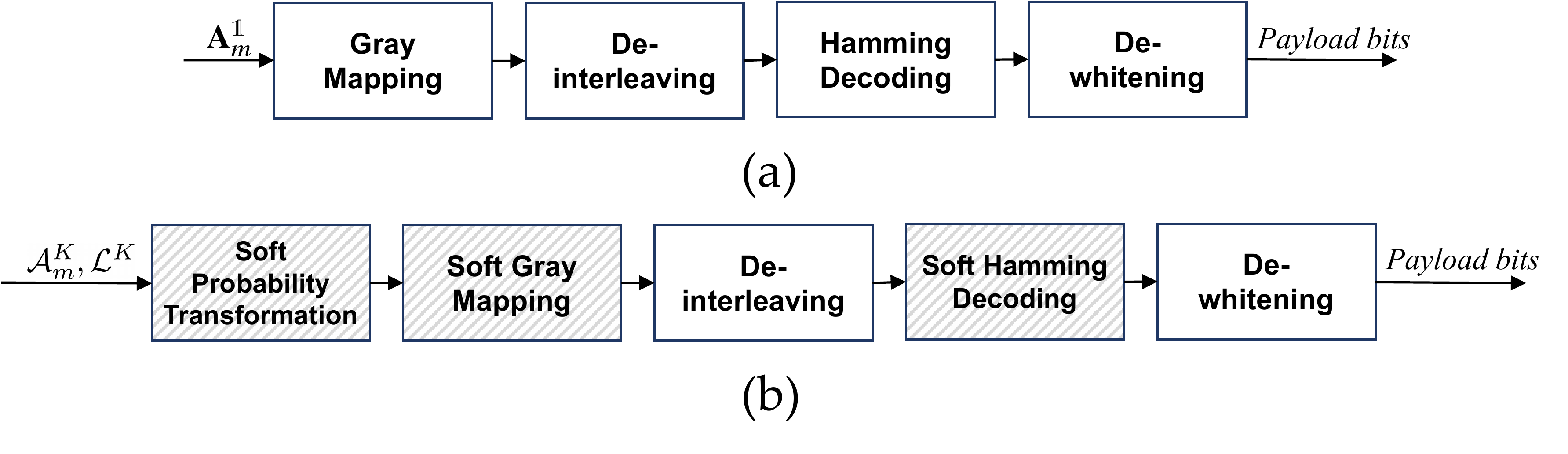}
    \caption{Building blocks and workflow of (a) standard LoRa hard channel decoder, and (b) soft channel decoder of LoRaPDA. Blocks in gray are new blocks in the soft decoder.}
    \label{fig:soft_arch}
\end{figure}

\begin{figure} [t]
	\setlength{\belowcaptionskip}{0.5cm}
	\vspace{-0.3cm}
	\centering
	\includegraphics[width=0.48\textwidth]{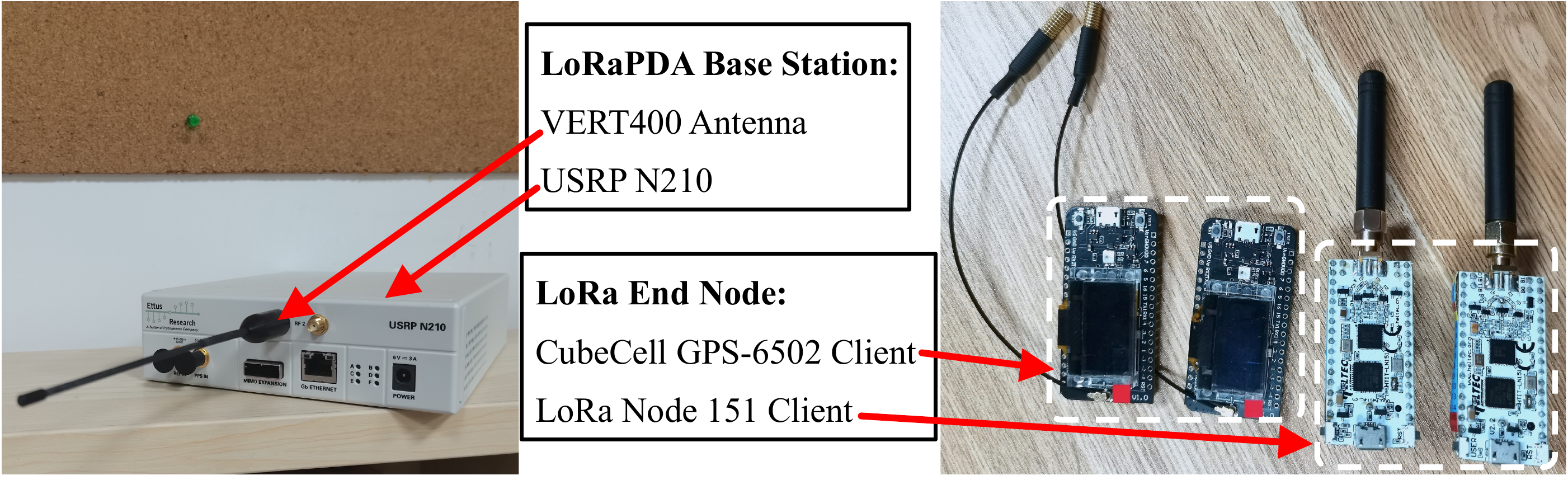}
	\caption{Testbed of LoRaPDA.}
	\label{fig:testbed}
	\vspace{-0.25cm}
\end{figure}

\subsection{Soft Channel Decoding} 

Fig.~\ref{fig:soft_arch} shows the building blocks and workflow of the conventional hard channel decoder and the soft channel decoder in LoRaPDA. The de-interleaver and de-whitening blocks remain the same, and the three gray blocks are newly designed. 
Below will elaborate on these new blocks.

\emph{1) Soft Probability Conversion:} Before performing soft gray mapping, we need to convert the top-$K$ per-symbol probabilities into per-bit soft probabilities as shown in Algorithm~\ref{algo:per_bit_pr}.
The target is to get probabilities of $SF$ bits $P_{s_m^{[i]}}=\{P_{d_{SF-1}},\cdots,P_{d_0}\}$ that are equal to zero (the probabilities that are equal to one can be computed accordingly), where $d_n$ indicates the $n$-th bit. We iterate over each bit in sequence $\mathcal{A}_m^K[i]$, and get corresponding indexes $idx_0$ and $idx_1$ that are equal to $0$ and $1$. With the indexes, we can separate and sum the log-likelihood probabilities for $0$ and $1$ on $\mathcal{L}^K[i]$. Then we normalize the soft probability for each bit. Note that the log-likelihood probabilities are all below zero, and closer-to-zero values mean higher probabilities. Thus, we put the summed log likelihood of ones $l_1$ in the numerator for normalized probabilities of zero.

\emph{2) Soft Gray Mapping:} %
To compute the normalized probability after gray mapping, we need to transform the probabilities of some bits jointly.
First, we need to transfer the probabilities of the -1 shift operation. Take -1 shift of three bits $(d_2, d_1, d_0)$ for example. The probability that $d_2=0$ can be extracted from (0, 0, 0), (0, 0, 1), (0, 1, 0), and (0, 1, 1), whose probabilities can be converted from (0, 0, 1), (0, 1, 0), (0, 1, 1), and (1, 0, 0), respectively. Specifically, the probability that $d_2$ is zero after the -1 shift is equal to the probability of two cases: (1) $d_2=1$, $d_1=d_0=0$, and (2) $d_2=0$, $d_1\&d_0\neq 0$.
Thus, the probability after -1 shift $P_{d_n}^{'}$ is given by

\vspace{-5mm}
\begin{equation}
	P_{d_n}^{'} = (1-P_{d_n}) \cdot \prod_{\theta=0}^{n-1} P_{d_\theta}
	+P_{d_n} \cdot (1-\prod_{\theta=0}^{n-1} P_{d_\theta}), 0< n<SF,
	\label{eq:soft_gray_pr_minus_1}
\end{equation}

\noindent and $P_{d_0}^{'}=1-P_{d_0}$. With $P_{d_n}^{'}$, we can then compute the probabilities for XOR-ing operations. 
Then the probability after XOR-ing $P_{d_n}^{''}$ can be written as

\vspace{-5mm}
\begin{equation}
	P_{d_n}^{''} = P_{d_n}^{'} \cdot P_{d_{n+1}}^{'}
	+(1-P_{d_n}^{'}) \cdot (1-P_{d_{n+1}}^{'}), 0\leq n<SF-1,
	\label{eq:soft_gray_pr_xor}
\end{equation}

\noindent and $P_{d_{SF-1}}^{'}=1$. Repeating the above steps for all bits, we can get the soft probabilities after gray mapping as
$\{P^{''}_{s_m^{(0)}},\cdots,P^{''}_{s_m^{(N_s-1)}}\}$. 

\begin{figure}[t]
	\setlength{\belowcaptionskip}{0.5cm}
	\vspace{-0.3cm}
	\centering
	\subfloat[]{%
		\includegraphics[width=0.5\columnwidth]{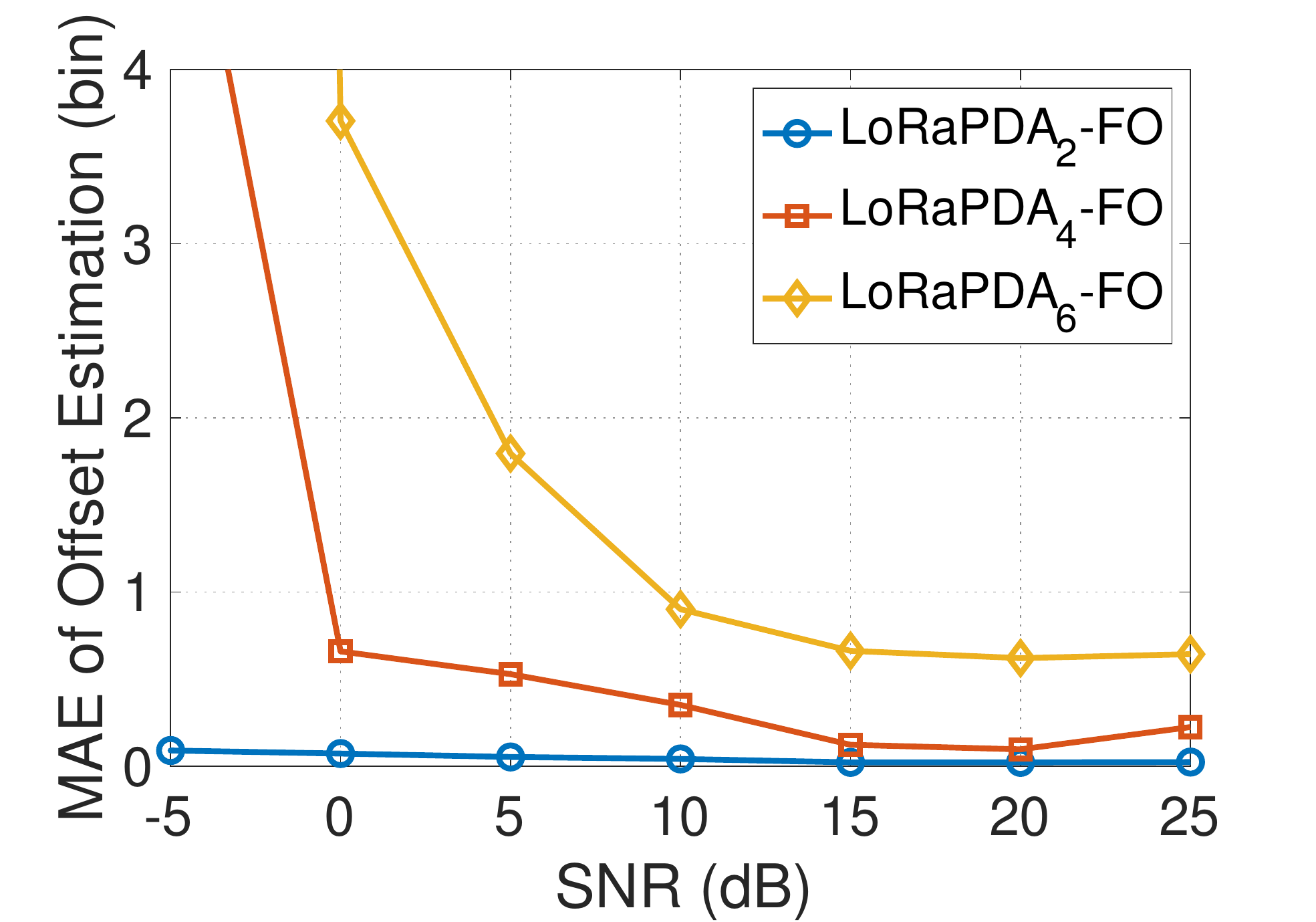}
		\vspace{-0.15cm}
		\label{fig:sim_pream_est_diff_num_user_mae}}
	\subfloat[]{%
		\includegraphics[width=0.5\columnwidth]{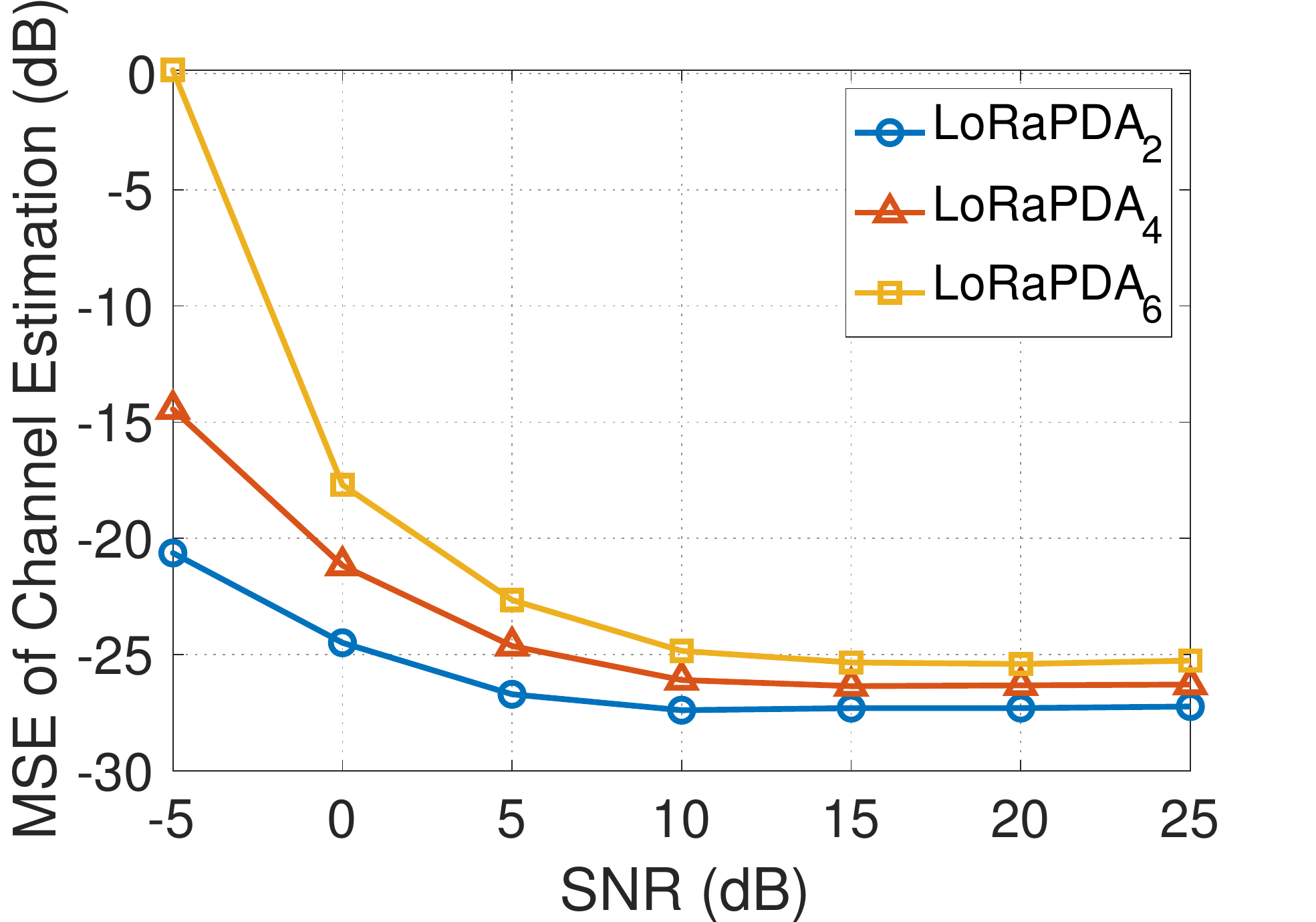}
		\vspace{-0.15cm}
		\label{fig:sim_pream_est_diff_num_user_mse}}
	\caption{Channel and offset estimation results for different number of concurrent transmissions: (a) MAE of offset estimation; (b) MSE of air-channel estimation.}
	\label{fig:sim_pream_est_diff_num_user}
\end{figure}

\emph{3) Soft Hamming Decoder:}  The Hamming block codes adopted by the standard LoRa decoder do not support probabilistic inputs. Thus, we adopt the syndrome-based soft decoder\cite{muller2011low} that has a linear computation complexity and improved performance of 1.3 dB compared with the traditional hard-decision Hamming decoder. It searches for possible error patterns which belong to the same syndrome and suggests the one with the highest probability of correct decoding.
Binary bit outputs of the soft Hamming decoder are passed to the de-whitening block, and the final payload bits are obtained.

\ignore{
    \begin{figure} [t]
    	\setlength{\belowcaptionskip}{0.5cm}
    	\vspace{-0.3cm}
    	\centering
    	\includegraphics[width=\columnwidth]{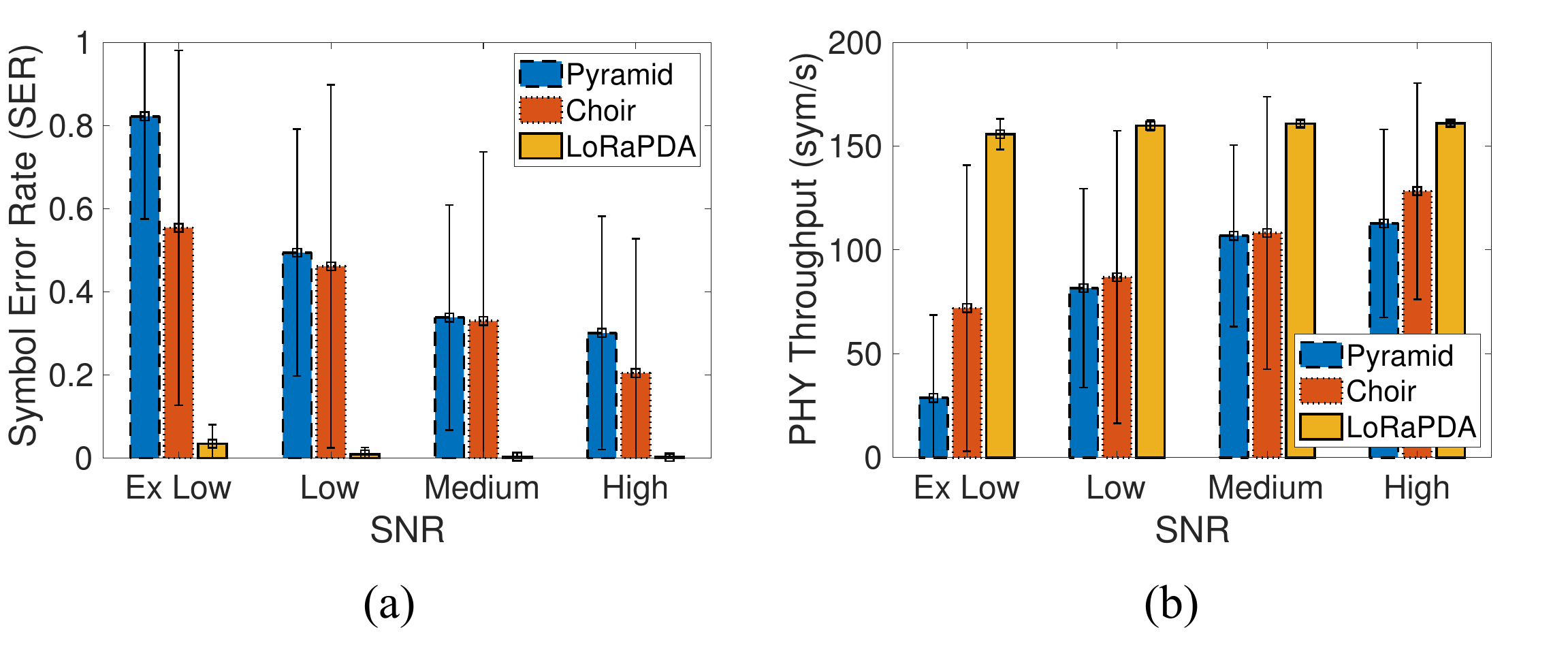}
    	\caption{Physical-layer performance comparison (two-user). (a) SER and (b) physical-layer throughput comparison among LoRaPDA and other MPR methods under small TO ($10\% \cdot T_s$).}
    	\label{fig:real_phy_diff_algo}
    	\vspace{-0.8cm}
    \end{figure}
}

\section{Implementation and Evaluation} \label{sec:exp}

\subsection{Implementation}%

\textbf{Hardware.} We have implemented a testbed using a USRP N210 and two types of certified COTS LoRa nodes designed by Heltec as shown in Fig.~\ref{fig:testbed}. USRP is used as a gateway equipped with a UBX daughterboard and a single VERT400 antenna. It can receive signals at 470 MHz bands, and support a maximal sampling rate of 20M samples per second.
LoRa Node 151 is equipped with an STM32 MCU and an SX1276 radio. CubeCell GPS-6502 is equipped with an ASR6502 MCU and an SX1262 radio. All nodes can transmit and receive signals at 470MHz - 510MHz with a single antenna. The center frequency of all LoRa nodes is set to 470 MHz.

\textbf{Software.} We use UHD to log LoRa baseband samples and implement LoRaPDA algorithms in MATLAB to demodulate and decode the samples offline. Each LoRa packet contains a 10-upchirp preamble, a 2.25-downchirp SFD, and a 12-Bytes payload with 24 symbols after encoding. By default, we use the spreading factor $SF=10$, coding rate $CR=\frac{4}{8}$, and bandwidth $BW=125$ KHz in our experiment. The symbol duration $T_s$ is around 8ms, and the packet duration $T_p$ is around 300ms. 
The sampling rate of LoRaPDA is set to 1M samples per second to allow oversampling. %
LoRaPDA is a gateway-side software solution that is compatible with COTS LoRa nodes without any hardware modifications on LoRa nodes.

\begin{figure}[t]
	\setlength{\belowcaptionskip}{0.5cm}
	\vspace{-0.3cm}
	\centering
	\subfloat[]{%
		\includegraphics[width=0.5\columnwidth]{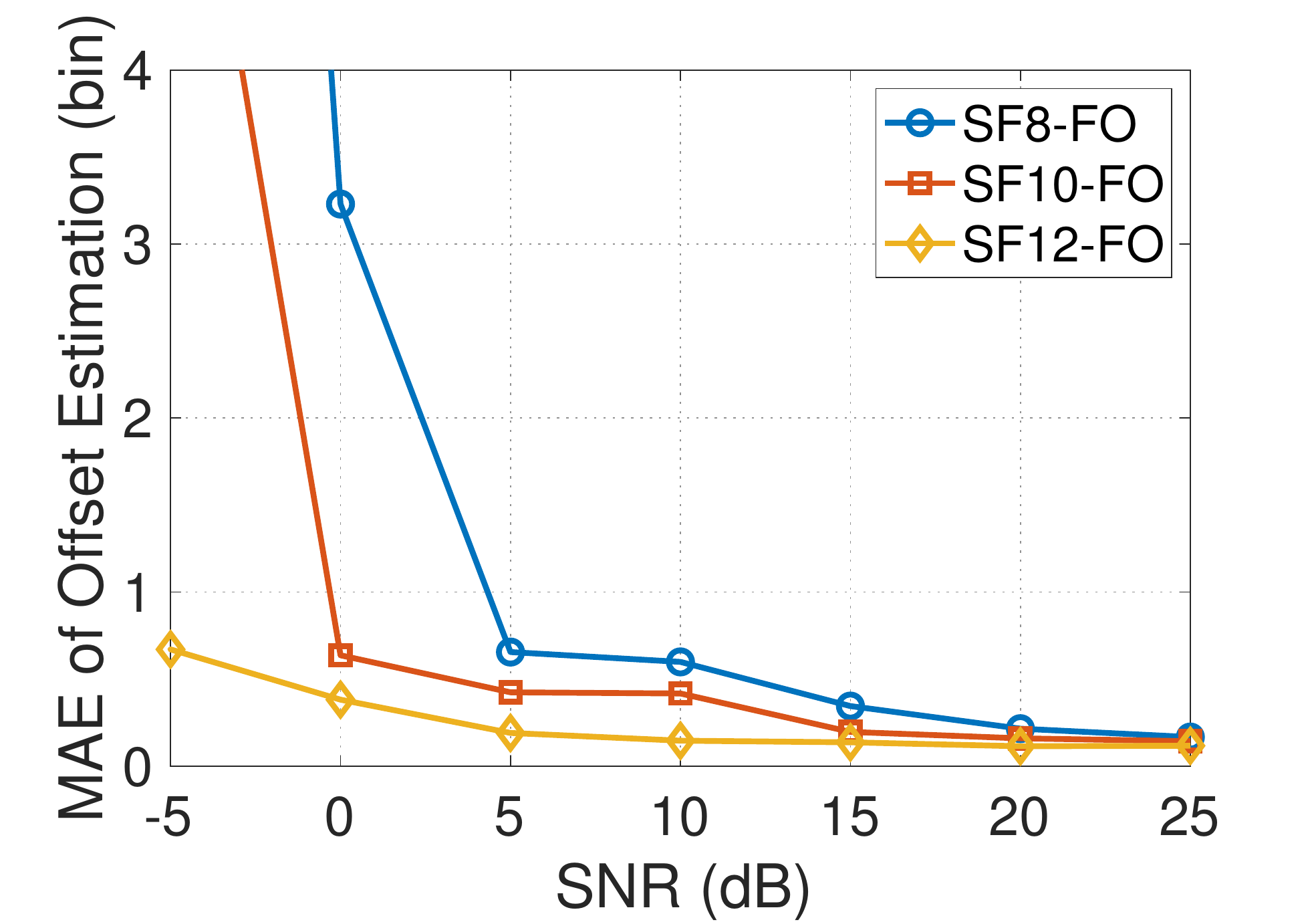}
		\vspace{-0.15cm}
		\label{fig:sim_pream_est_diff_sf_mae}}
	\subfloat[]{%
		\includegraphics[width=0.5\columnwidth]{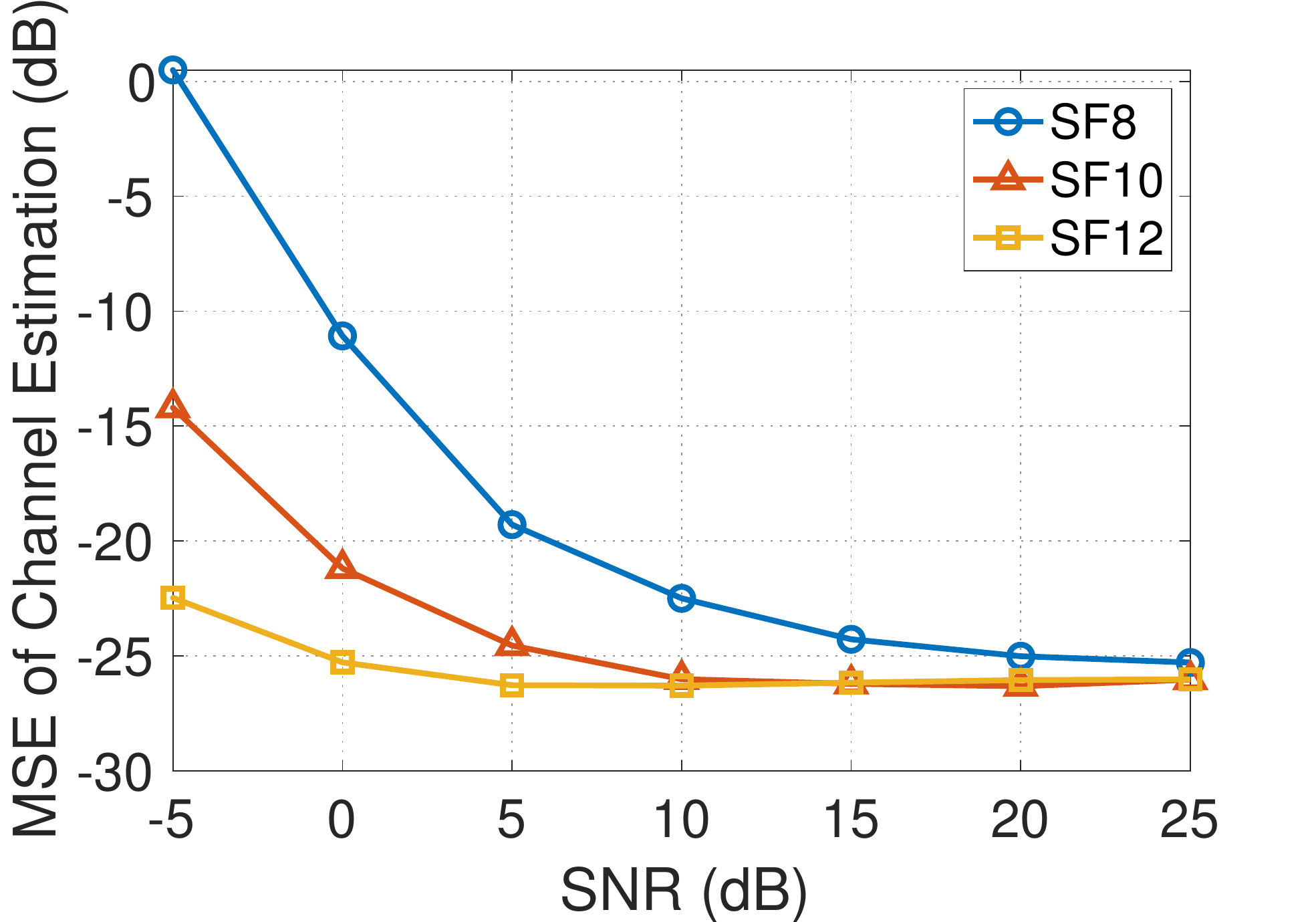}
		\vspace{-0.15cm}
		\label{fig:sim_pream_est_diff_sf_mse}}
	\caption{Channel and offset estimation results for different SFs with four-user concurrent transmissions: (a) MAE of offset estimation; (b) MSE of air-channel estimation.}
	\label{fig:sim_pream_est_diff_sf}
\end{figure}

\subsection{Experiment Setup}
\textbf{Methodology.} 
We generate 2000 concurrent transmissions and plot average results to compare the performance with other MPR methods. To better control TOs, we conduct \emph{trace-driven} simulations to evaluate the performance of LoRaPDA. Specifically, we collect baseband samples from different nodes at high SNR using a USRP N210 in an indoor environment. We slightly change the center frequency of USRP to introduce random CFOs (i.e., 470MHz+$\epsilon$KHz) so that the overall CFOs of logged packets are within [-5KHz, 5KHz].
We then align collected packet traces with random TOs to emulate concurrent transmissions (the maximal TO is $10\% \cdot T_s$). 
Power difference between any two packets is larger than 1dB so that their preambles and SFDs are differentiable for channel and offset estimation.
Finally, we add noise to the superimposed signal to accurately emulate different SNRs, i.e., high SNR (\textgreater15dB), medium SNR (5$\sim$15dB), low SNR (-5$\sim$5dB), and extremely low SNR (\textless-5dB).

\ignore{
	\textbf{Configurations.} We evaluate the performance of our algorithms by using the following configurations.
	\begin{itemize}
		\item \emph{SF and BW:} we use the configurations of SF=7 and BW=250KHz.
		\item \emph{Payload Length:} we use payload length of 20 Bytes, leading to 38 payload chirps after channel coding.
		\item \emph{Packet Structure:} we use implicit header mode in the end node for aggregation application. The remaining structure consists of preamble, sync word, payload, and CRC.
	\end{itemize}
}

\textbf{Algorithms.} We only compare LoRaPDA with MPR solutions that are compatible with COTS nodes. We consider two LoRaPDA decoders: 
\begin{itemize}
	\item \emph{LoRaPDA-Hard}: The hard decoder  that uses the standard Hamming decoder with hard demodulation results in LoRaPDA;
	\item \emph{LoRaPDA-Soft}: The soft decoder  that uses the design in Section \ref{sec:design:soft} with soft-demodulation results in LoRaPDA.
\end{itemize}

\noindent We compare LoRaPDA with the following MPR algorithms:

\begin{itemize}
	\item Pyramid\cite{xu2021pyramid}: It is a typical approach that represents sliding-window-based decoding approaches. We adopt its open-source implementation to handle collided signals under small TOs. %
	\item Choir\cite{eletrebychoir17}: It is an approach that may work under small TOs with a single antenna. We re-implement the algorithm following the Choir paper.
	The reason why we do not compare with PCube \cite{xia2021pcube} is that PCube is a multiple-antenna solution, while LoRaPDA is a single-antenna solution.
\end{itemize}

\textbf{Metrics.} Since we use MPR to decode each user's data for aggregation, we reuse the following MPR metrics, which also represent the performance of data aggregation:
\begin{itemize}
	\item \emph{Symbol Error Rate (SER)} is defined as the percentage of erroneous symbols in received symbols. Note that SER only applies to \emph{LoRaPDA-Hard}.
	\item \emph{Physical-layer (PHY) Throughput} is defined as the number of correct symbols versus transmission time in symbol per second (sym/s). Similarly, it only applies to \emph{LoRaPDA-Hard}.
	\item \emph{Bit Error Rate (BER)} is defined as the percentage of erroneous bits in received bits after channel decoding.
	\item \emph{Network-layer (NET) Throughput} is defined as the number of correct bits that pass the CRC check versus transmission time in bit per second (bit/s). It will take re-transmissions into account, and use a simple MAC protocol defined in Section~\ref{sec:exp:results}.
\end{itemize}

\begin{figure*}[t]
	\setlength{\belowcaptionskip}{0.5cm}
	\centering
	\subfloat[]{%
		\includegraphics[width=0.25\textwidth]{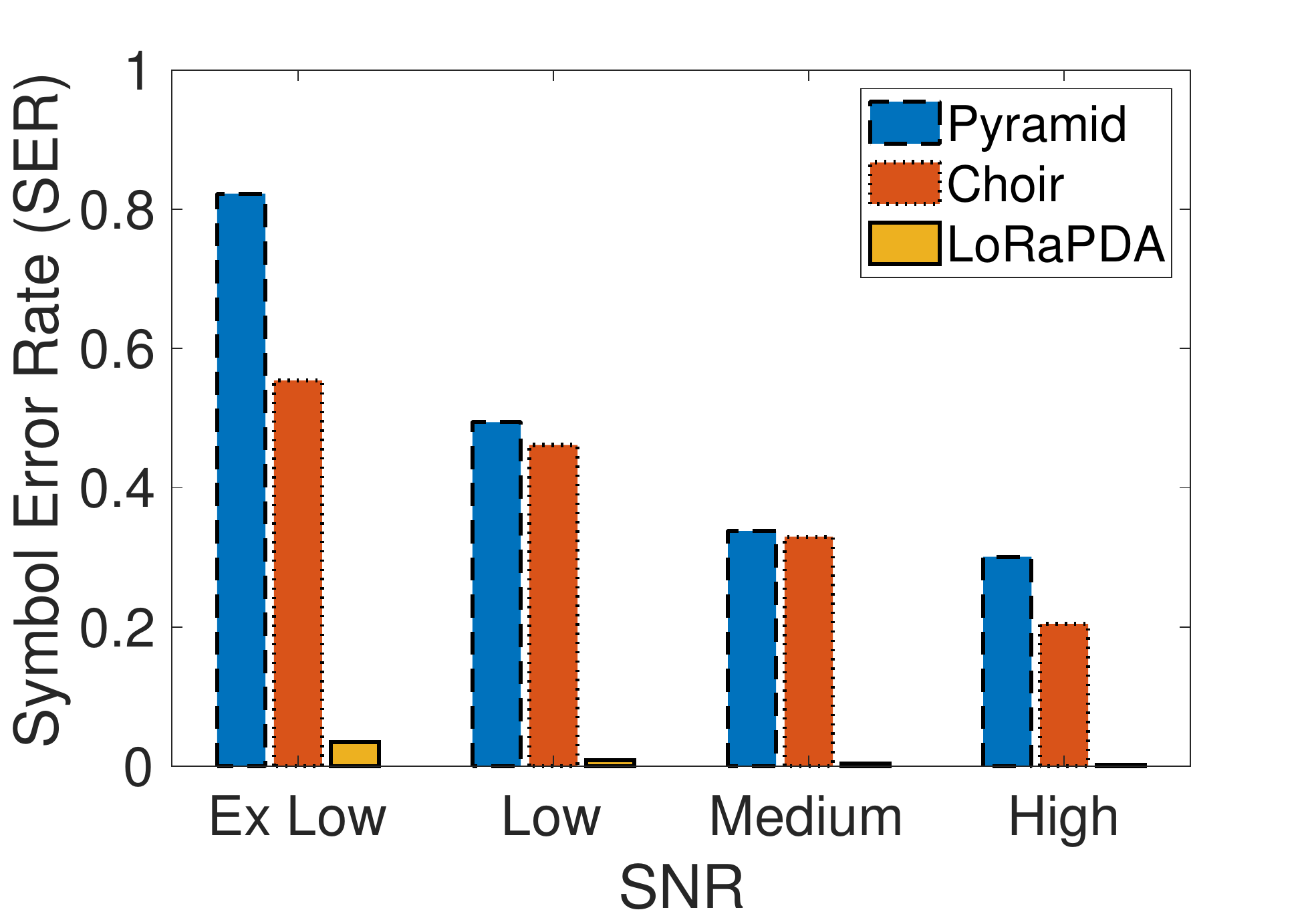}
		\label{fig:real_phy_diff_algo_ser_2U}}
	\subfloat[]{%
		\includegraphics[width=0.25\textwidth]{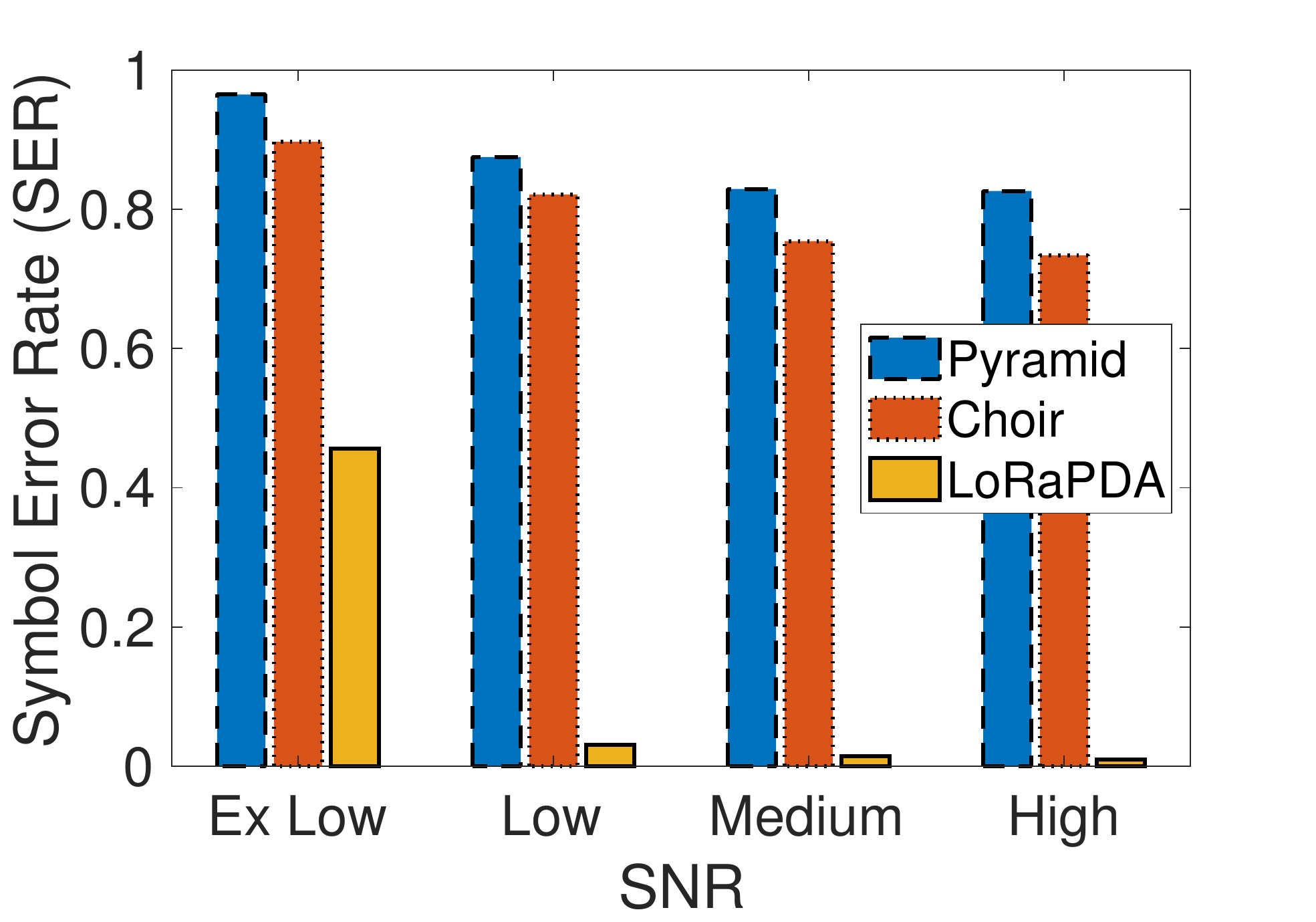}
		\label{fig:real_phy_diff_algo_ser_4U}}
	\subfloat[]{%
		\includegraphics[width=0.25\textwidth]{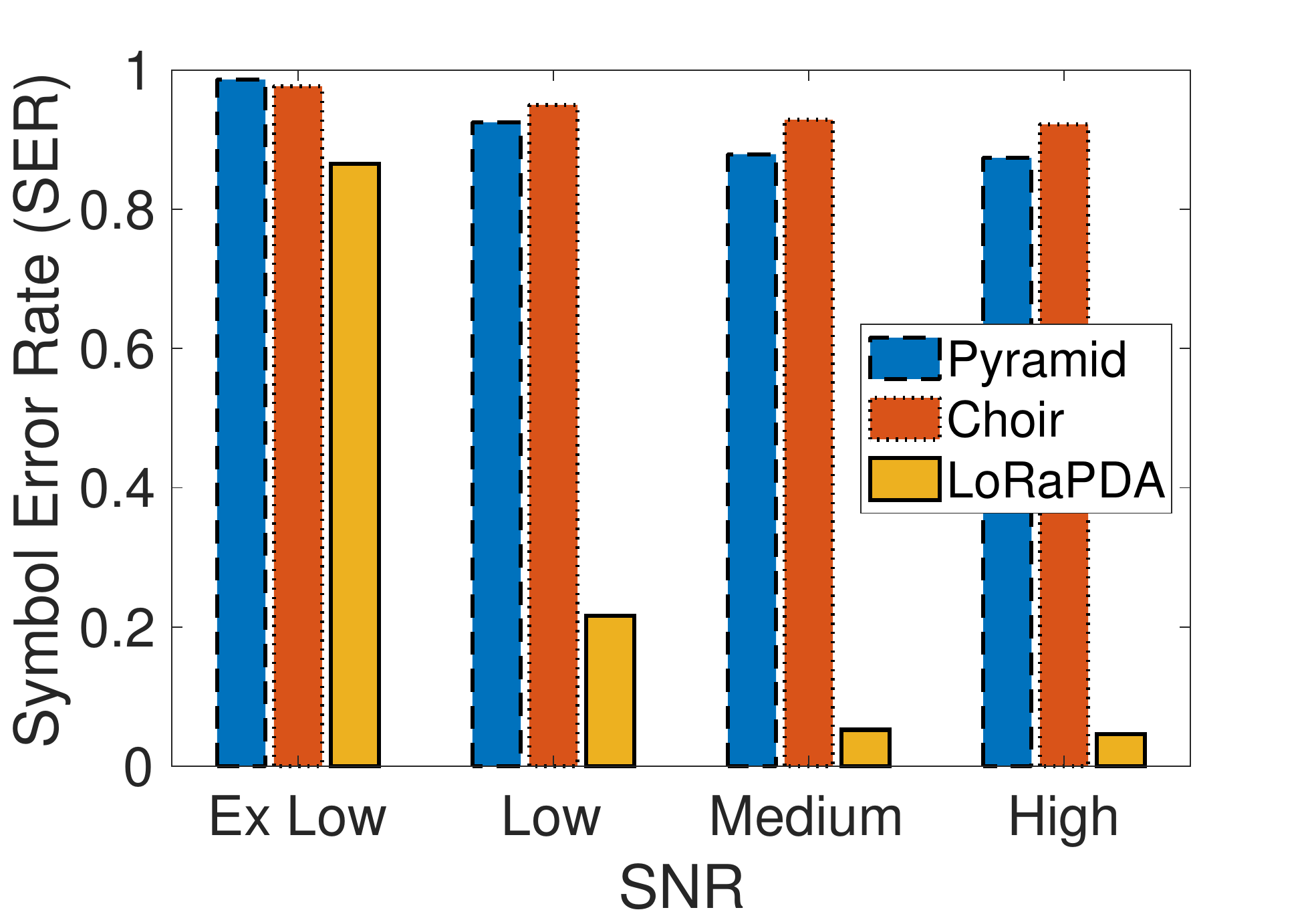}
		\label{fig:real_phy_diff_algo_ser_6U}}
	\subfloat[]{%
		\includegraphics[width=0.25\textwidth]{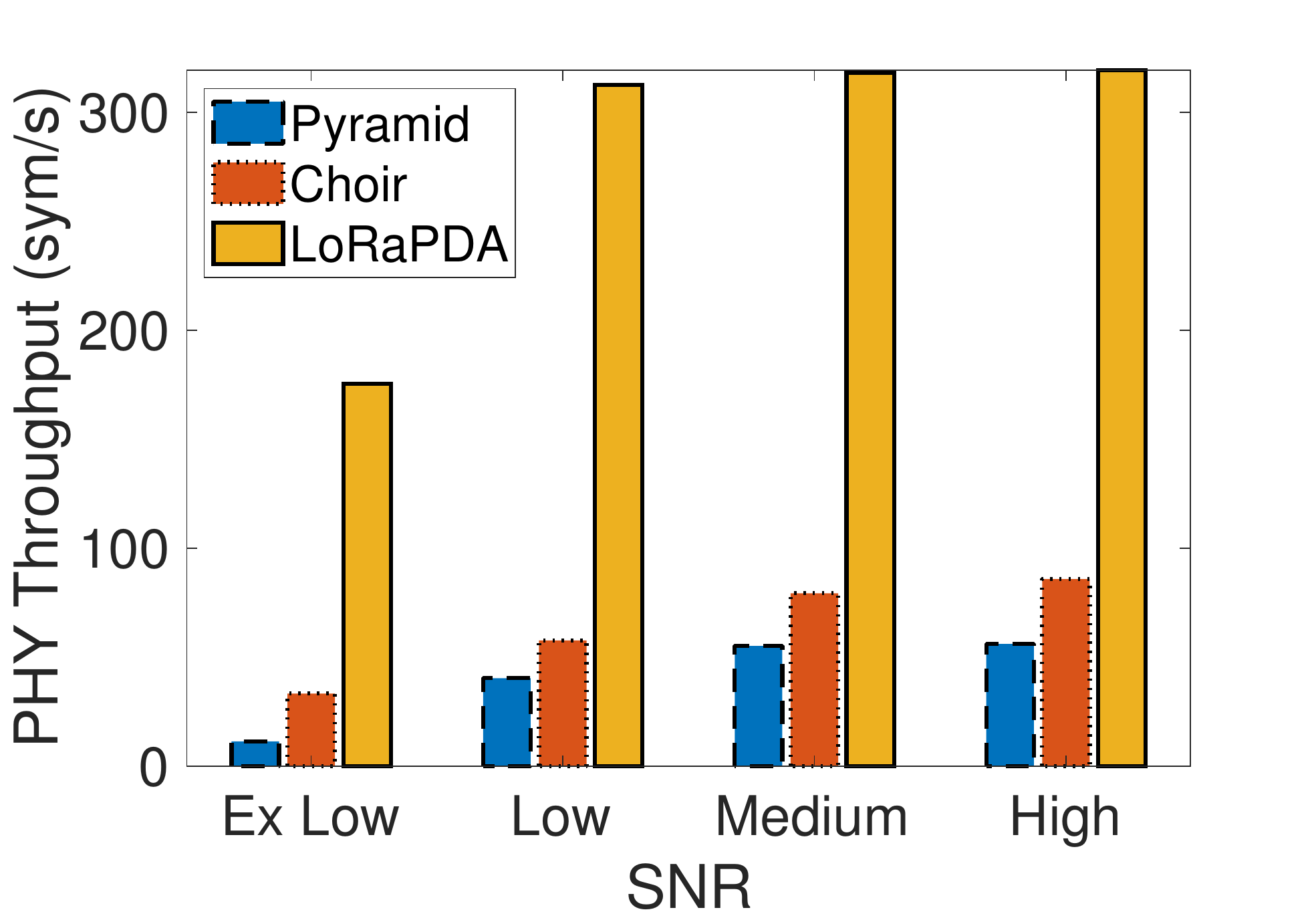}
		\label{fig:real_phy_diff_algo_thrpt_4U}}
	\caption{Symbol error rate between LoRaPDA and other MPR methods under $SF=10$ with (a) two-user concurrent transmissions, (b) four-user concurrent transmissions, and (c) six-user concurrent transmissions; (d) physical-layer throughput between LoRaPDA and other MPR methods under $SF=10$ with four-user concurrent transmissions.}
	\label{fig:real_phy_diff_algo}
	\vspace{-0.2cm}
\end{figure*} 

\subsection{Physical-layer Performance} \label{sec:exp:results}

\emph{1) \textbf{Channel and Offset Estimation.}} We adopt the simulation approach that the ground truth is known to evaluate the performance of channel and offset estimation.
We first study the performance under different number of concurrent transmissions for LoRaPDA. %
Fig.~\ref{fig:sim_pream_est_diff_num_user} shows the channel and offset estimation results under two, four and six-user concurrent transmissions. %
As we can see,
the MAE of frequency offset under four-user concurrent transmissions is comparable with two-user concurrent transmissions for SNR$\geq$0dB. Moreover, the MSE of channel estimation under four-user concurrent transmissions is still comparable, although its performance degrades due to a higher probability of co-located preambles. However, for six-user concurrent transmissions, the FO estimation accuracy drops significantly when SNR$\leq$10dB, and the air-channel estimation accuracy also drops at low SNRs. This is due to even higher co-located preamble probability for more concurrent transmissions. %

Then we study the impact of $SF$ on the channel and offset estimation accuracy assuming four-user concurrent transmissions. Fig.~\ref{fig:sim_pream_est_diff_sf} shows that $SF=10$ achieves almost the same performance as $SF=12$ when SNR$\geq$0dB, and is better than $SF=8$ in both channel and offset estimation. The phenomenon can also be explained by the lower collision probability when $SF$ is larger. Given that channel and offset estimation is crucial for demodulation and decoding, the performance under $SF=8$ may not be acceptable. Thus, we adopt $SF=10$ for LoRaPDA to balance the preamble estimation accuracy and physical-layer data rate.

\emph{2) \textbf{Symbol Demodulation.}} %
We compare SER of LoRaPDA with that of Pyramid and Choir with different number of concurrent transmissions under different SNR regimes for $SF=10$.
As shown in Fig.~\ref{fig:sim_colocated}, two-user, four-user, and six-user concurrent transmissions have different co-located transmission probabilities. Thus, the comparison results can show the MPR performance of LoRaPDA under different levels of co-located peaks.

Fig.~\ref{fig:real_phy_diff_algo}(a)
shows that SER of LoRaPDA is much lower than that of other MPR methods under two-user concurrent transmissions. When SNR is extremely low, the SER of Pyramid and Choir reaches $82.2\%$, and $55.4\%$, respectively, while that of LoRaPDA is only $3.4\%$. When the SNR is high, the SER of LoRaPDA is even lower than $0.2\%$, while the SER of Pyramid and Choir remains higher than $20\%$, meaning that both Pyramid and Choir do not work well under small TOs even with few co-located peaks.

Fig.~\ref{fig:real_phy_diff_algo}(b)
compares their performance under four-user concurrent transmissions with more co-located peaks. The performance of LoRaPDA degrades slightly for low to high SNRs except for the extremely low SNR where the performance can be easily affected by noise. However, its performance is significantly better than Pyramid and Choir, e.g., 5.3$\times$ of Choir (175.5 sym/s versus 33.3 sym/s) and 15.4$\times$ of Pyramid (175.5 sym/s versus 11.4 sym/s) for low SNR as shown in
Fig.~\ref{fig:real_phy_diff_algo}(d).
Fig.~\ref{fig:real_phy_diff_algo}(c)
shows a similar trend for six-user concurrent transmissions, although LoRaPDA's performance further degrades at low SNR. 

In summary, LoRaPDA outperforms Choir and Pyramid under small TOs, from few co-located peaks to many co-located peaks. Choir's channel estimation accuracy can be easily affected by noise, leading to performance degradation.
Pyramid's degraded performance is due to no distinct TO between packets, and the apex patterns can not be separated. LoRaPDA, however, can work under small TOs and remains robust to co-located peaks.

\begin{figure*}[t]
	\setlength{\belowcaptionskip}{0.5cm}
	\centering
	\begin{minipage}[t]{0.245\textwidth}
		\centering
		\includegraphics[width=\textwidth]{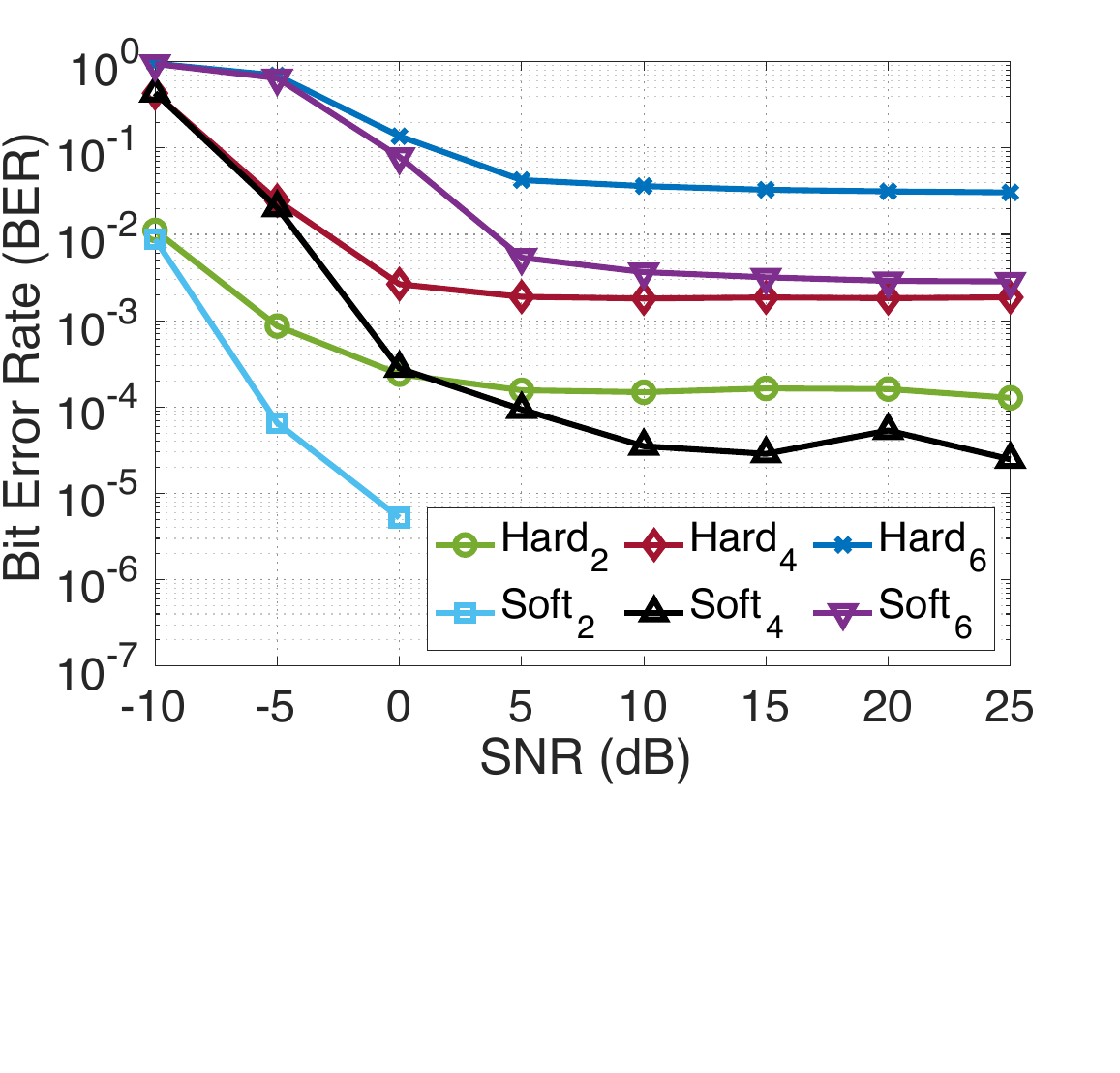}
		\caption{BER of LoRaPDA-Hard and LoRaPDA-Soft.}
		\label{fig:real_phy_hard_soft_ber}
	\end{minipage}
	\hfill
	\begin{minipage}[t]{0.49\textwidth}
		\centering
		\includegraphics[width=\textwidth]{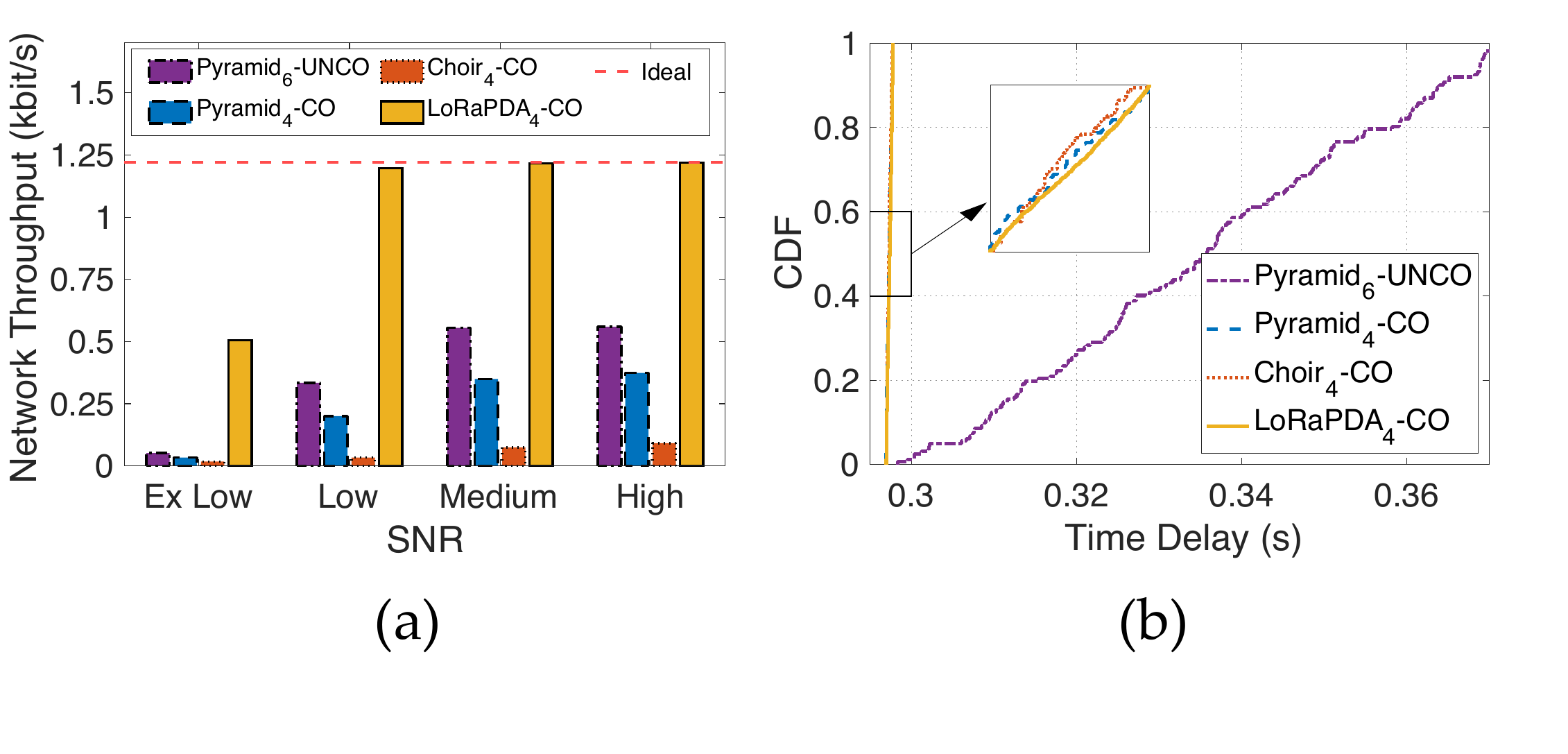}
		\caption{(a) Network throughput of different transmission schemes; (b) CDF of packet collection delay at high SNR.}
		\label{fig:real_netwk}
	\end{minipage}
        \hfill
        \begin{minipage}[t]{0.245\textwidth}
		\centering
		\includegraphics[width=\textwidth]{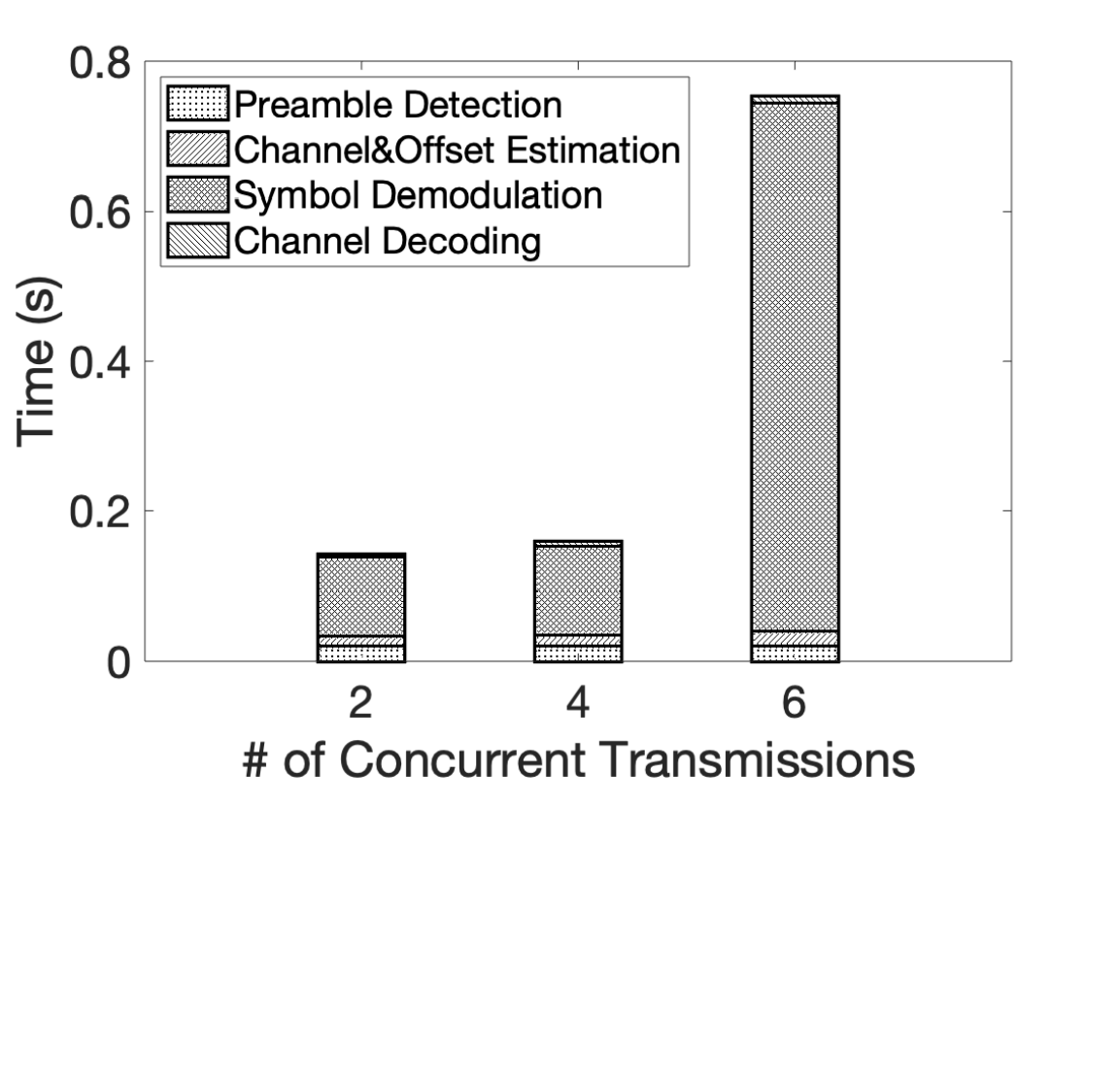}
		\caption{Real-time decoding performance.}
		\label{fig:real_time_perf}
	\end{minipage}
	\vspace{-0.25cm}
\end{figure*}

\emph{3) \textbf{Channel Decoding.}} We compare the decoding performance of \emph{LoRaPDA-Soft} and \emph{LoRaPDA-Hard} with two-user, four-user and six-user concurrent transmissions under $SF=10$. 
The subscript number indicates the number of users.
We keep top-$2$ likelihood probabilities in symbol demodulation for \emph{LoRaPDA-Soft}.

Fig.~\ref{fig:real_phy_hard_soft_ber} shows the average BER under -10$\sim$25dB SNR. For two-user decoding, the BER of LoRaPDA-Hard$_2$ has a floor above $1\times10^{-4}$ even at high SNR, while the BER of LoRaPDA-Soft$_2$ reaches below $1\times10^{-4}$ when SNR$\approx$-5 dB, and even achieves below $1\times10^{-6}$ when SNR\textgreater 0dB. For four-user decoding, LoRaPDA-Hard$_4$ has a floor above $1\times10^{-3}$, while LoRaPDA-Soft$_4$ can still achieve below $1\times10^{-4}$ when SNR$\geq$5dB. For SNR$\geq$0dB, the BER of LoRaPDA-Soft is 10$\times$ lower than LoRaPDA-Hard. This is because LoRaPDA-Soft keeps close-to-maximal sequences and likelihood probabilities, which improves the decoding performance. Moreover, for six-user decoding, both LoRaPDA-Hard$_6$ and LoRaPDA-Soft$_6$ have higher error floors ($10^{-1}$ and $10^{-2}$), due to lower channel estimation and offset accuracy. Thus, we use \emph{LoRaPDA-Soft} and four-user concurrent transmissions in the following network-layer simulation.

\ignore{
\begin{figure}[t]
	\setlength{\belowcaptionskip}{0.5cm}
	\vspace{-0.3cm}
	\centering
	\subfloat[]{%
		\includegraphics[width=0.5\columnwidth]{figures/fig_real_diff_sf_ser.pdf}
		\vspace{-0.15cm}
		\label{fig:real_diff_sf_ser}}
	\hspace{-5mm}
	\subfloat[]{%
		\includegraphics[width=0.5\columnwidth]{figures/fig_real_diff_to_ser.pdf}
		\vspace{-0.15cm}
		\label{fig:real_diff_to_ser}}
	\caption{Impact of LoRa parameters. SER performance with different (a) SFs and (b) TOs under different SNRs.}
	\label{fig:real_parameter_impacts}
	\vspace{-0.8cm}
\end{figure}
}

\subsection{Network-layer Performance}
We further compare the network-layer performance of LoRaPDA and other MPR methods with coordinated (CO) and uncoordinated (UNCO) transmissions. %
Packets are transmitted in rounds and are without re-transmission.
For UNCO transmission, the maximal TO among packets is set to 0.25$T_p$ ($\sim$75ms), and a random delay is added to different users using the ALOHA MAC protocol. %
For CO transmission, we set the maximal TO to 0.1$T_s$ ($\sim$1ms). %
We define the packet data collection delay as the time between the round start time and the time that a packet is received correctly. We use the same signal-level simulation approach for network-layer simulations. Only packets that pass the CRC check are delivered to the network layer. 
We use the time of 300 seconds to evaluate the NET throughput. %

Fig.~\ref{fig:real_netwk}(a) shows the NET throughput results under different SNR regimes. Denote six-user concurrent transmissions of Pyramid as Pyramid$_6$ (it is reported to support at most six users in \cite{xu2021pyramid}), and four-user concurrent transmissions of Choir and LoRaPDA as Choir$_4$ and LoRaPDA$_4$, respectively. Compared with the UNCO transmissions, the NET throughput of LoRaPDA$_4$-CO is at least 2.1$\times$ of Pyramid$_6$-UNCO at all SNRs (e.g., 1.2 kbit/s vs. 0.56 kbit/s at high SNR). This is due to two reasons: 1) small symbol TOs (two symbols are almost aligned though their packet TO are large) still occur under $SF=10$ when the number of users increases, leading to demodulation errors in Pyramid, and 2) UNCO transmissions cost more time than CO transmissions. 
Compared with other CO transmissions, the NET throughput of LoRaPDA$_4$-CO is 3.2$\times$ of Pyramid$_4$-CO (0.37 kbit/s), and 13.3$\times$ of Choir$_4$-CO (0.09 kbit/s) at high SNR. The main reason is that BER of LoRaPDA under small TOs is lower. Moreover, LoRaPDA$_4$-CO achieves close-to-optimal throughput (ideal=1.22 kbit/s) for low, medium and high SNRs.

Fig.~\ref{fig:real_netwk}(b) shows the data collection delay CDF of CO and UNCO transmission at high SNR. The delay of CO transmission is within 0.3 seconds (around a packet duration). The delay of UNCO transmission is longer than that of CO transmission due to the ALOHA protocol, which also explains the lower throughput of Pyramid$_6$-UNCO. Although the delay of CO transmissions is similar, Choir is not robust to noise, and Pyramid can only decode less than two packets on average, leading to lower NET throughput than LoRaPDA. Overall, compared with other MPR-based physical-layer data aggregation solutions, LoRaPDA has the potential for quick data aggregation queries.

\subsection{Real-time Decoding Performance} \label{sec:exp:real-time}

We implement a real-time LoRaPDA system with two USRPs. One USRP serves as the concurrent transmission LoRa nodes, and the other USRP serves as the receiver. To ease implementation, we directly generate superimposed signals of concurrent transmissions in Matlab, and use USRP to transmit the baseband signal. The receiver is implemented on a Linux server with Intel Xeon Gold 5122 CPUs, 64GB DDR4 DRAM, and an Nvidia RTX 2080 Ti GPU. 
Note that it is possible to use a server for LoRa signal processing, where all baseband samples are transferred to the cloud, as many advanced LoRa systems \cite{dongare2018charm,liu2020nephalai} have adopted the same decoding model for improved performance.

To ease the interaction with GPU, we implement all receiver blocks in Python. Most blocks are implemented in the CPU, and only the symbol demodulation block is implemented in GPU with CUDA version 11.4. For the CPU implementation, we only use two threads: one thread receives a block of samples from USRP and puts them into a queue, and the other thread extracts the samples from the queue for further sequential processing. %
We have verified the receiver's correctness by comparing its decoding result with the Matlab implementation.

Fig.~\ref{fig:real_time_perf} shows the decoding time of each block under two-user, four-user and six-user concurrent transmissions.
As we can see, the decoding time of two-user and four-user is less than the packet duration ($\sim$0.3s), while the decoding time of six-user is less than three times the packet duration. Although the decoding time of six-user exceeds the packet duration, it is still comparable to the packet duration given that there are six concurrent transmissions. Moreover, the symbol demodulation block is the most complex block, where the demodulation time increases a lot for six-user since its enumeration sequence increases a lot. However, thanks to the GPU hardware, we manage to implement a real-time decoder.

\section{Related Work} \label{sec:related}

\textbf{Physical-layer Data Aggregation (PDA)}: 
QuAiL \cite{gadre2020quick} also targets PDA in LPWAN, but it adopts an analog computation approach, which is sensitive to phase misalignment common in practice and is not compatible with COTS LoRa nodes. CompAir \cite{abari2016over} proposes a precoding protocol to first align transmitted signals' phases, and then uses the analog computation approach for PDA. The severe protocol overhead may not be feasible for low-cost LoRa nodes. 

Physical-layer data aggregation is closely related to the over-the-air computation (AirComp) concept in the wireless community. Refs. \cite{gastpar2003source, nazer2007computation, goldenbaum2013harnessing} prove that the AirComp rate surpasses that of the orthogonal transmission scheme. 
Ref. \cite{Goldenbaum2009airfunction} shows that the non-linear aggregation function can be implemented through summation with some simple pre-processing and post-processing. 
Recently, ref. \cite{zhu2019broadband,zhu2020one,yang2020federated} considers OFDM-based AirComp for federated edge learning. However, they all assume perfect phase alignment and use the analog computation approach. Refs. \cite{huang2020physical} and \cite{zhao2022broadband} uses an digital computation approach for AirComp on OFDM under practical phase-asynchronous channel. Instead, we target digital AirComp on LoRa CSS.

\textbf{MPR in Wireless Systems}: MPR is similar to collision decoding in wireless systems. ZigZag \cite{gollakota2008zigzag} combats the hidden terminal problem in 802.11 WiFi networks with the binary phase shift keying (BPSK) modulation. Ref. \cite{halperin2008taking} and mZig \cite{kong2015mzig} realize collision decoding in 802.15.4 ZigBee networks with the offset quadrature phase shift keying (OQPSK) modulation. BiGroup \cite{ou2015come} resolves the tag collisions in RFID networks with the on-off keying (OOK) modulation. In contrast, LoRaPDA targets collision decoding in LoRa with the CSS modulation, which is different from BPSK, OQPSK and OOK.

LoRa collision decoding is widely studied in recent years. Most methods that are compatible with COTS LoRa nodes are introduced in Section~\ref{sec:mov:sota}. There are some other methods that modify the LoRa physical layer for MPR. FlipLoRa \cite{xu2020fliplora} encodes data with interleaved upchirp and downchirp to help decode collided packets. CurvingLoRa \cite{li2022curvinglora} encodes data with non-linear chirps so that tiny TOs lead to negligible interference, and the MPR performance is improved. These work is orthogonal to LoRaPDA: LoRaPDA can also use these new modulation methods to further improve MPR performance.

\section{Conclusion} \label{sec:conclusion}
This paper presents LoRaPDA, a novel LoRa system that leverages the MPR technique to compute aggregate data from multi-users almost time-synchronized but phase-misaligned superimposed signal. In particular, LoRaPDA leverages a new symbol demodulation algorithm using maximum likelihood detection, and a new soft-decision packet decoding algorithm with improved decoding performance.
Experiment results show that LoRaPDA supports up to six concurrent transmissions with low SER, and outperforms the SOTA MPR algorithm by 5.3$\times$ and 2.1$\times$ in terms of physical-layer and network-layer throughput, respectively. In the future, we are interested in a) studying new channel and offset estimation approach that supports more concurrent transmissions, and b) implementing a complete real-time system with time-synchronized LoRa nodes.

\appendices

\ifCLASSOPTIONcaptionsoff
  \newpage
\fi

\bibliographystyle{IEEEtran}
\bibliography{ref}

\end{document}